\documentclass[twoside]{article}
\usepackage{amsmath,amssymb,times}
\usepackage{graphicx}
\usepackage[T1]{fontenc}
\usepackage[french]{babel}
\usepackage{amsthm}
\usepackage{amsfonts}
\usepackage{amssymb, amscd}
\usepackage{epsfig,color}
\usepackage{fancyhdr}
\def\shorttitle{\sc Relativit\'e g\'en\'erale}
\leftmark
\def\shortauthor{\sc\leftmark}

\input epsf

\def\lb{\label}
\def\bb{\bibitem}
\def\be{\begin{equation}}
\def\ee{\end{equation}}
\def\ba{\begin{eqnarray}}
\def\ea{\end{eqnarray}}
\def\ds{\displaystyle}
\def\nn{\nonumber}

\def\ol{\overline}
\def\e{{\rm e}}
\def\A{{\cal A}} \def\F{{\cal F}} \def\T{{\cal T}} \def\M{{\cal M}}
\def\E{{\cal E}}
\def\x{\vec{x}}

\begin{document}
\date{}
\title{
\begin{flushright}\begin{small}    LAPTH-Conf-030/10
\end{small} \end{flushright} \vspace{2cm}
Relativit\'e g\'en\'erale: solutions exactes stationnaires}

\author{\textbf{G\'erard Cl\'ement}}

\maketitle
\begin{abstract}
Apr\`es des rappels sur les fondements de la relativit\'e
g\'en\'erale, nous nous int\'eresserons aux solutions exactes
stationnaires des \'equations d'Einstein et d'Einstein-Maxwell. Un
certain nombre de ces solutions peuvent \^etre interpr\'et\'ees
comme des trous noirs, dont nous \'etudierons la g\'eom\'etrie
d'espace-temps. Puis nous verrons comment la reformulation des
\'equations d'Einstein-Maxwell stationnaires comme un mod\`ele sigma
autogravitant fournit un outil puissant pour la g\'en\'eration de
nouvelles solutions.
\end{abstract}

\begin{center}
{\textbf{Abstract}} \vspace{0.5cm}

\begin{minipage}{10cm}
After a brief summary of the foundations of general relativity, we
will concentrate on the stationary exact solutions of the Einstein
and Einstein-Maxwell equations. A number of these solutions can be
interpreted as black holes, corresponding to a special spacetime
geometry. Then, we will show that the reformulation of the
stationary Einstein-Maxwell equations as a gravitating sigma model
provides us with a powerful tool to generate new solutions.

\end{minipage}
\end{center}
\newpage
$ $
\newpage

\tableofcontents

\newpage
\section*{Introduction}
La relativit\'e g\'en\'erale d'Einstein est l'une des th\'eories
physiques les plus \'el\'egantes. L'id\'ee fondamentale \`a la base
de cette th\'eorie est que le champ de gravitation, que nous
mesurons par ses effets sur le mouvement des corps massifs, est en
fait une manifestation de la g\'eom\'etrie courbe de l'espace-temps.
Apr\`es sa formulation il y a bient\^ot un si\`ecle, ses
pr\'edictions \`a l'\'echelle astrophysique (d\'eviation de la
lumi\`ere par les corps massifs, pr\'ecession du p\'erih\'elie des
plan\`etes, d\'ecalage des raies spectrales vers le rouge) ont
\'et\'e rapidement confirm\'ees par l'observation. Une pr\'ediction
moins \'evidente de la th\'eorie, qui a mis beaucoup plus longtemps
\`a se d\'egager, est maintenant aussi bien confirm\'ee par les
observations. C'est celle de l'existence des trous noirs, des
r\'egions de l'espace-temps o\`u la courbure est si forte que la
lumi\`ere -- ainsi que tous les objets tomb\'es dans ces r\'egions
-- est pi\'eg\'ee et ne peut plus s'en \'echapper.

La nature de la g\'eom\'etrie spatio-temporelle des trous noirs peut
\^etre \'elucid\'ee par l'\'etude de certaines solutions exactes des
\'equations d'Einstein, ou des \'equations coupl\'ees
d'Einstein-Maxwell. Ces solutions stationnaires (ind\'ependantes du
temps), dont la plus connue est la solution de Schwarzschild, feront
l'objet de ce cours.

Dans une premi\`ere partie, nous rappellerons bri\`evement les
notions de base de la relativit\'e g\'en\'erale qui seront
utilis\'ees dans la suite du cours. La seconde partie sera
consacr\'ee \`a l'\'etude de la g\'eom\'etrie des trous noirs
stationnaires, en commen\c{c}ant par la solution de Schwarzschild.
Nous verrons comment la singularit\'e apparente de cette solution
sur la sph\`ere de rayon $r = 2GM$ peut \^etre \'elimin\'ee par la
transformation de coordonn\'ees de Kruskal, cette sph\`ere
(l'horizon \'ev\`enement) correspondant simplement \`a une
fronti\`ere entre l'ext\'erieur du trou noir, o\`u la g\'eom\'etrie
spatio-temporelle est qualitativement la m\^eme que celle de
l'espace-temps de Minkowski, et l'int\'erieur du trou noir, o\`u le
c\^one de lumi\`ere a bascul\'e de sorte que les r\^oles du temps et
de la coordonn\'ee radiale sont \'echang\'es. Nous introduirons
ensuite les diagrammes de Penrose, qui permettent de visualiser
simplement la structure causale des espace-temps stationnaires. Puis
nous verrons que les \'equations d'Einstein-Maxwell admettent
\'egalement une solution trou noir, la solution de
Reissner-Nordstr\"{o}m, dont la structure plus complexe (existence
de deux horizons) peut \^etre ais\'ement visualis\'ee \`a l'aide de
diagrammes de Penrose. Apr\`es ces solutions \`a sym\'etrie
sph\'erique, nous \'etudierons la solution de Kerr, qui est
seulement \`a sym\'etrie axiale, et repr\'esente le champ
gravitationnel \`a l'ext\'erieur d'un objet massif en rotation
autour de son axe de sym\'etrie. Enfin, nous mentionnerons tr\`es
rapidement quelques \'el\'ements de la thermodynamique des trous
noirs.

La troisi\`eme partie sera ax\'ee sur la g\'eom\'etrie, non plus de
l'espace-temps, mais de l'``hyperespace'' des solutions
stationnaires des \'equations d'Einstein ou d'Einstein-Maxwell. Nous
verrons comment ces solutions peuvent \^etre reli\'ees entre elles
par des transformations appartenant \`a un groupe de sym\'etrie
cach\'e du syst\`eme d'\'equations des champs. Nous montrerons
d'abord comment r\'eduire les \'equations d'Einstein stationnaires
\`a celles d'une th\'eorie effective de gravitation \`a seulement
trois dimensions, avec comme source un champ dit de ``mod\`ele
sigma'' associ\'e au groupe $SL(2,R)$. Puis nous passerons au cas
des \'equations d'Einstein-Maxwell stationnaires, dont le groupe
d'invariance est $SU(2,1)$, et verrons comment une transformation du
groupe permet de g\'en\'erer la solution de Reissner-Nordstr\"{o}m
\`a partir de celle de Schwarzschild. Nous verrons ensuite comment
la combinaison de transformations du groupe $SU(2,1)$ avec une
transformation de coordonn\'ees permet de g\'en\'erer la solution de
Kerr \`a partir de celle de Schwarzschild. Enfin, nous ferons une
excursion vers la th\'eorie d'Einstein pour l'espace-temps \`a cinq
dimensions, dont l'espace des solutions 2-stationnaires (ne
d\'ependant pas de deux des cinq coordonn\'ees) admet le groupe
d'invariance $SL(3,R)$, et montrerons comment l'action de ce groupe
permet de relier entre elles diverses solutions de cette th\'eorie.

Le temps limit\'e nous a oblig\'e \`a omettre d'autres aspects de la
th\'eorie et de la physique des trous noirs. Mentionnons les trous
noirs en gravitation \`a trois dimensions, qui ont \'et\'e
\'etudi\'es en particulier par une collaboration
alg\'ero-fran\c{c}aise \cite{tmgbh,adtmg,tmgebh}. La dynamique des
trous noirs, qui fait l'objet de nombreux travaux num\'eriques:
collisions de trous noirs, et formation de trous noirs par
effondrement gravitationnel. Et les aspects quantiques, avec entre
autres la possibilit\'e de production de micro-trous noirs par des
collisionneurs de particules.

Parmi les nombreux livres sur les sujets abord\'es ici, citons
\cite{ll,weinberg,mtw,wald,kenyon} pour la relativit\'e
g\'en\'erale, \cite{hawell,wald,chandra,haw,exact} pour la th\'eorie
des trous noirs, et \cite{exact} pour les mod\`eles sigma.

\section{Rappels de relativit\'e g\'en\'erale}
\setcounter{equation}{0}

La relativit\'e g\'en\'erale est une th\'eorie du champ de
gravitation. Comme pour toute th\'eorie des champs, on proc\'edera
pour la d\'efinir en deux \'etapes. On commencera d'abord par
d\'efinir le champ par son {\em effet} sur le mouvement d'une
particule d'\'epreuve, puis on cherchera la {\em cause} de ce champ,
pour arriver \`a formuler les \'equations r\'egissant le champ
produit par une source mat\'erielle. D'o\`u une division naturelle
en deux parties: principe d'\'equivalence, puis \'equations
d'Einstein.

\subsection{Principe d'\'equivalence}
\subsubsection{M\'ecanique non-relativiste}
En m\'ecanique non-relativiste, le mouvement d'une particule
d'\'epreuve isol\'ee (loin des autres corps) est r\'egi par la loi
d'inertie de Galil\'ee \be\lb{gal} \frac{d(m\vec{v})}{dt} = 0 \qquad
\left(\vec{v} \equiv \frac{d\vec{x}}{dt}\right)\,, \ee qui traduit
la conservation de la quantit\'e de mouvement $m\vec{v}(t)$, o\`u
$m$ est la {\em masse inerte} de la particule. En cons\'equence, la
particule d'\'epreuve est anim\'ee d'un mouvement rectiligne et
uniforme.

En pr\'esence d'autres corps massifs engendrant un champ de
gravitation $\vec{g}(\vec{x})$, le mouvement de la particule
d'\'epreuve est acc\'el\'er\'e suivant la loi de Newton \be
\frac{d(m\vec{v})}{dt} = m\vec{g}(\vec{x})\,. \ee La constante de
couplage $m$ de la particule au champ de gravitation figurant dans
le membre de droite de cette \'equation est la {\em masse
gravitationnelle} de la particule. L'exp\'erience montre que (en
l'absence de frottement) tous les corps d'\'epreuve, quelle que soit
leur masse inerte, sont acc\'el\'er\'es de la m\^eme fa\c{c}on par
un champ de gravitation, ce que traduit le principe d'\'equivalence,
qui affirme l'\'egalit\'e (dans un syst\`eme d'unit\'es convenable)
entre masse gravitationnelle et masse inerte.

\subsubsection{Relativit\'e restreinte}
En relativit\'e restreinte, la loi d'inertie prend la forme
covariante \be \frac{d(mu^{\mu})}{d\tau} = 0 \ee ($\mu = 0,1,2,3$),
o\`u le quadrivecteur vitesse \be u^{\mu} = \frac{dx^{\mu}}{d\tau}
\ee est la d\'eriv\'ee par rapport au temps propre $\tau$ du
quadrivecteur position $x^{\mu}$. Le temps propre est reli\'e au
temps coordonn\'ee $x^0 \equiv t$ par \be d\tau^2 =
\left(1-\frac{\vec{v}^2}{c^2}\right)dt^2\,, \ee $c$ \'etant la
vitesse de la lumi\`ere. Cette d\'efinition peut \^etre r\'e\'ecrite
sous forme covariante (en unit\'es $c=1$) \be - d\tau^2 = ds^2 =
-dt^2 + d\vec{x}^2 = \eta_{\mu\nu}dx^{\mu}dx^{\nu}\,, \ee o\`u $ds$
est l'intervalle infinit\'esimal, $\eta_{\mu\nu} =$
diag($-1,+1,+1,+1$) est le tenseur m\'etrique de Minkowski, et on a
utilis\'e la convention de sommation d'Einstein (la somme sur tous
les indices r\'ep\'et\'es une fois en haut et une fois en bas est
sous-entendue).

La loi d'inertie relativiste d\'erive du principe d'action
\ba\lb{act1} I
&=& -m\int d\tau = -m\int \sqrt{1-\frac{\vec{v}^2}{c^2}}\,dt \nn \\
&=& m\int \eta_{\mu\nu}u^{\mu}u^{\nu}d\tau \ea (toujours en unit\'es
$c=1$).

Si l'on passe des coordonn\'ees minkowskiennes $x^{\mu}$ \`a des
coordonn\'ees curvilignes quelconques $\hat{x}^{\mu}$ par la
transformation $$x^{\mu} \to \hat{x}^{\mu}\,,$$ l'intervalle
infinit\'esimal prend la forme \be ds^2 =
\eta_{\mu\nu}\frac{\partial x^{\mu}}{\partial\hat{x}^{\alpha}}
\frac{\partial x^{\nu}}{\partial\hat{x}^{\beta}}
d\hat{x}^{\alpha}d\hat{x}^{\beta} =
g_{\alpha\beta}d\hat{x}^{\alpha}d\hat{x}^{\beta}\,, \ee o\`u
$g_{\alpha\beta}$ est le tenseur m\'etrique en coordonn\'ees
curvilignes. L'int\'egrale d'action (\ref{act1}) prend la forme \be
I = m\int g_{\mu\nu}\hat{x}^{\mu}\hat{x}^{\nu}\,d\tau\,. \ee La
variation de cette action conduit, apr\`es int\'egration par
parties, \`a \ba \delta I &=& m\int \left[\partial_{\rho}g_{\mu\nu}
\hat{u}^{\mu}\hat{u}^{\nu} - 2\frac{d}{d\tau}(g_{\rho\nu}
\hat{u}^{\nu})\right]\delta\hat{x}^{\rho}d\tau \nn \\
&=& m\int \left[(\partial_{\rho}g_{\mu\nu} -
2\partial_{\mu}g_{\rho\nu})\hat{u}^{\mu}\hat{u}^{\nu} -
2g_{\rho\nu}\frac{d\hat{u}^{\nu}}{d\tau}\right]\delta\hat{x}^{\rho}d\tau \nn \\
&=& -2m\int g_{\rho\lambda}\frac{D\hat{u}^{\lambda}}{D\tau}
\delta\hat{x}^{\rho}d\tau\,, \ea o\`u on a utilis\'e $d/d\tau =
u^{\mu}\partial_{\mu}$, et d\'efini la d\'eriv\'ee covariante \be
\frac{D\hat{u}^{\lambda}}{D\tau} \equiv
\frac{d\hat{u}^{\lambda}}{d\tau} + \Gamma_{\mu\nu}^{\lambda}
\hat{u}^{\mu}\hat{u}^{\nu}\,, \ee o\`u les \be\lb{christ}
\Gamma_{\mu\nu}^{\lambda} \equiv
\frac12g^{\lambda\rho}(\partial_{\mu}g_{\rho\nu} +
\partial_{\nu}g_{\rho\mu} - \partial_{\rho}g_{\mu\nu})
\ee sont les symboles de Christoffel (sym\'etriques en $(\mu,\nu)$).
Dans (\ref{christ}), $g^{\lambda\rho}$ est le tenseur m\'etrique
contravariant, dont les composantes sont les \'el\'ements de la
matrice inverse de celle des $g_{\mu\nu}$:
$$g^{\lambda\rho}g_{\sigma\rho} = \delta_{\lambda}^{\sigma}$$
($\delta_{\lambda}^{\sigma}$ \'etant le symbole de Kronecker). La
loi d'inertie relativiste se traduit donc en coordonn\'ees
curvilignes par les \'equations de mouvement covariantes \be
\frac{D\hat{u}^{\lambda}}{D\tau} = 0\,, \ee

\subsubsection{Relativit\'e g\'en\'erale}
En relativit\'e g\'en\'erale, l'espace-temps de Minkowski est
remplac\'e par une vari\'et\'e riemanienne (\`a signature
lorentzienne) \`a quatre dimensions, munie de la m\'etrique \be ds^2
= g_{\mu\nu}(x)dx^{\mu}dx^{\nu}\,, \ee o\`u le tenseur m\'etrique
$g_{\mu\nu}(x)$ (matrice sym\'etrique inversible de signature
$-+++$) peut \^etre diagonalis\'e (ramen\'e \`a $\eta_{\mu\nu}$)
localement par une transformation de coordonn\'ees, mais pas
globalement.

La {\em d\'eriv\'ee covariante} d'un quadrivecteur de composantes
contravarantes $A^{\mu}(x)$ est d\'efinie par \be\lb{dercov1}
D_{\nu}A^{\mu} =
\partial_{\nu}A^{\mu} + \Gamma_{\nu\rho}^{\mu}A^{\rho}\,, \ee o\`u
les connexions $\Gamma_{\nu\rho}^{\mu}$ sont donn\'ees par les
symboles de Christoffel (\ref{christ}). La d\'eriv\'ee covariante
(\ref{dercov1}) se transforme comme un tenseur mixte. De m\^eme, la
d\'eriv\'ee covariante de la 1-forme de composantes covariantes \be
A_{\mu} = g_{\mu\nu}A^{\nu} \quad \leftrightarrow \quad  A^{\mu} =
g^{\mu\nu}A_{\nu} \ee est d\'efinie par \be\lb{dercov2}
D_{\nu}A_{\mu} = \partial_{\nu}A_{\mu} -
\Gamma_{\mu\nu}^{\rho}A_{\rho}\,, \ee Ces d\'efinitions se
g\'en\'eralisent facilement aux cas de d\'eriv\'ees covariantes de
tenseurs contravariants, covariants ou mixtes. En particulier, la
d\'eriv\'ee covariante du tenseur m\'etrique \be\lb{dercovg}
D_{\lambda}g_{\mu\nu} = \partial_{\lambda}g_{\mu\nu} -
\Gamma_{\mu\lambda}^{\rho}g_{\rho\nu} -
\Gamma_{\nu\lambda}^{\rho}g_{\rho\mu} = 0 \ee s'annulle, en vertu de
la d\'efinition (\ref{christ}) de la connexion. Inversement, la
condition (\ref{dercovg}) entra\^{\i}ne l'\'egalit\'e entre la
connexion et le symbole de Christoffel (\ref{christ}).

Le postulat de la relativit\'e g\'en\'erale d'Einstein est que les
particules d'\'epreuve non isol\'ees ob\'eissent toujours au
principe d'action (\ref{act1}) qui conduit (comme dans le cas de
l'espace-temps de Minkowski param\'etris\'e par des coordonn\'ees
curvilignes) aux \'equations \be\lb{geo} \frac{D{u}^{\mu}}{D\tau}
\equiv u^{\nu}D_{\nu}u^{\mu} = \frac{d{u}^{\mu}}{d\tau} +
\Gamma_{\nu\rho}^{\mu} {u}^{\nu}{u}^{\rho} = 0\,, \ee Ces
\'equations expriment que les particules d'\'epreuve suivent des
{\em g\'eod\'esiques} de genre temps de la vari\'et\'e riemannienne.
On montre que ces \'equations admettent l'int\'egrale premi\`ere \be
g_ {\mu\nu}u^{\mu}u^{\nu} = -\alpha^2\ee ($\alpha^2$ constante
r\'eelle), \'equivalente (moyennant une red\'efinition de l'unit\'e
de temps propre) \`a la d\'efinition $d\tau^2=-ds^2$ du temps
propre.

Les \'equations (\ref{geo}) peuvent aussi \^etre \'ecrites sous la
forme quasi-newtonienne \be \frac{d(mu^{\mu})}{d\tau} =
-m\Gamma_{\nu\rho}^{\mu}(x) {u}^{\nu}{u}^{\rho}\,, \ee o\`u le
membre de droite repr\'esente la ``quadriforce'' relativiste
agissant sur la particule. Dans la limite non-relativiste ($u^i \to
v^i$, $u^0 \to 1$, $\tau \to t$), ces \'equations se r\'eduisent \`a
\be \frac{d(mv^{i})}{dt} \simeq -m\Gamma_{00}^{i}(x) \,. \ee Le
terme $-\Gamma_{00}^{i}(x)$ peut donc \^etre interpr\'et\'e comme le
champ de gravitation newtonien agissant sur la particule. A
l'approximation des champs faibles ($g_ {\mu\nu} \simeq
\eta_{\mu\nu}$) lentement variables, ce champ se r\'eduit \`a \be
g^i = - \Gamma_{00}^{i}\simeq \frac12\,\partial_ig_{00}\,. \ee Il en
r\'esulte que, moyennant ces approximations, la composante $g_{00}$
du tenseur m\'etrique est reli\'ee au potentiel de gravitation
newtonien par \be g_{00} \simeq -(1 + 2U)\,. \ee

\subsection{Equations d'Einstein}
\subsubsection{Tenseur de Riemann}
Les connexions $\Gamma_{\nu\rho}^{\mu}(x)$ mesurent un champ de
gravitation apparent, qui peut toujours \^etre annul\'e localement
par une transformation de coordonn\'ees. Comment mesurer un vrai
champ de gravitation (qui ne peut pas \^etre annul\'e partout)?

La r\'eponse est fournie par la consid\'eration de la {\em
d\'eviation g\'eod\'esique} entre deux particules d'\'epreuve qui
suivent deux g\'eod\'esiques voisines $\Gamma$ et $\Gamma +
\delta\Gamma$. La g\'eod\'esique $\Gamma$ est d\'efinie par \be
\frac{D{u}^{\mu}}{D\tau} = u^{\nu}D_{\nu}u^{\mu} = 0\,. \ee La
g\'eod\'esique voisine ob\'eit \`a la m\^eme \'equation o\`u le
point d'espace-temps $x^{\mu}$ est remplac\'e par $x^{\mu} + \delta
x^{\mu}$. Au premier ordre, la diff\'erence entre les \'equations
des deux g\'eod\'esiques est donc \ba\lb{devgeo} 0 &=&
\delta(u^{\nu}D_{\nu}u^{\mu}) = \delta
x^{\lambda}D_{\lambda}(u^{\nu}D_{\nu}u^{\mu}) \nn\\
&=& \left[\delta x^{\lambda}D_{\lambda},u^{\nu}D_{\nu}\right]u^{\mu}
+ u^{\nu}D_{\nu}(\delta x^{\lambda}D_{\lambda}u^{\mu}) \nn\\
&=& \delta x^{\lambda}u^{\nu}\left[D_{\lambda},D_{\nu}\right]u^{\mu}
+ u^{\nu}D_{\nu}(u^{\lambda}D_{\lambda}\delta x^{\mu}) \ea (pour
passer de la deuxi\`eme \`a la troisi\`eme ligne, on a utilis\'e
$\delta x^{\lambda}D_{\lambda}u^{\mu} = \delta u^{\mu} = D\delta
x^{\mu}/D\tau$).

D\'efinissons le tenseur de courbure, ou tenseur de Riemann,
$R^{\mu}_{\rho\lambda\nu}$ par: \be\lb{defrie}
\left[D_{\lambda},D_{\nu}\right]A^{\mu} \equiv
{R^{\mu}}_{\rho\lambda\nu}A^{\rho}\,. \ee L'\'equation
(\ref{devgeo}) prend la forme ``newtonienne'' reliant une
acc\'el\'eration \`a une force \be \frac{D^2\delta x^{\mu}}{D\tau^2}
= {R^{\mu}}_{\rho\nu\lambda}u^{\rho}u^{\nu}\delta x^{\lambda}\,. \ee
En utilisant pour le calcul de (\ref{defrie}) les d\'efinitions des
d\'eriv\'ees covariantes d'un vecteur et d'un tenseur mixte, on
obtient les composantes du tenseur de Riemann \be
{R^{\mu}}_{\rho\nu\lambda} =
\partial_{\nu}\Gamma_{\rho\lambda}^{\mu} -
\partial_{\lambda}\Gamma_{\rho\nu}^{\mu} +
\Gamma_{\sigma\nu}^{\mu}\Gamma_{\rho\lambda}^{\sigma} -
\Gamma_{\sigma\lambda}^{\mu}\Gamma_{\rho\nu}^{\sigma}\,. \ee

Le tenseur de Riemann satisfait aux propri\'et\'es de sym\'etrie \ba
{R^{\mu}}_{\rho\nu\lambda} &=& - {R^{\mu}}_{\rho\lambda\nu}\,, \nn \\
{R^{\mu}}_{\rho\nu\lambda} + {R^{\mu}}_{\nu\lambda\rho} +
{R^{\mu}}_{\lambda\rho\nu} &=& 0\,, \nn \\ R_{\mu\rho\nu\lambda} &=&
R_{\nu\lambda\mu\rho} \ea o\`u $R_{\mu\rho\nu\lambda} \equiv
g_{\mu\sigma}{R^{\sigma}}_{\rho\nu\lambda}$. En vertu de ces
relations, le tenseur de Riemann a seulement 20 composantes
lin\'eairement ind\'ependantes (dans le cas de l'espace-temps \`a
quatre dimensions). Le tenseur de Riemann satisfait \'egalement aux
{\em identit\'es de Bianchi} (qui jouent pour la gravitation
d'Einstein le m\^eme r\^ole que le premier groupe d'\'equations de
Maxwell pour l'\'electromagn\'etisme) \be\lb{bianchi1}
D_{\sigma}{R^{\mu}}_{\rho\nu\lambda} +
D_{\lambda}{R^{\mu}}_{\rho\sigma\nu} +
D_{\nu}{R^{\mu}}_{\rho\lambda\sigma} = 0\,. \ee

En contractant le premier et le troisi\`eme indices du tenseur de
Riemann (en identifiant ces indices et en sommant dessus), on
obtient le tenseur de Ricci \be R_{\mu\nu} \equiv
{R^{\lambda}}_{\mu\lambda\nu}\,. \ee Ce tenseur est sym\'etrique,
donc a (toujours \`a quatre dimensions) 10 composantes
lin\'eairement ind\'ependantes. La contraction du tenseur de Ricci
conduit au scalaire de courbure \be R \equiv g^{\mu\nu}R_{\mu\nu}\,.
\ee Enfin, la contraction des identit\'es de Bianchi
(\ref{bianchi1}) conduit aux identit\'es (appel\'ees \'egalement
identit\'es de Bianchi) \be\lb{bianchi2}
D_{\nu}\left({R_{\mu}}^{\nu} - \frac12R\delta_{\mu}^{\nu}\right) =
0\,. \ee

\subsubsection{Equations d'Einstein}
Les identit\'es de Bianchi (\ref{bianchi2}) sont formellement
analogues \`a l'\'equation de continuit\'e covariante pour le
tenseur d'impulsion-\'energie $T_{\mu\nu}$ de la mati\`ere. En
relativit\'e restreinte, ces \'equations de continuit\'e (qui
rappellent l'\'equation de continuit\'e $\partial_{\nu}J^{\nu} = 0$
pour le courant \'electromagn\'etique) s'\'ecrivent dans un rep\`ere
minkowskien $\partial_{\nu}{T_{\mu}}^{\nu} = 0$, et leur
g\'en\'eralisation covariante en relativit\'e g\'en\'erale est
\be\lb{cont} D_{\nu}{T_{\mu}}^{\nu} = 0\,. \ee

Cette analogie entre (\ref{bianchi2}) et (\ref{cont}) a conduit
Einstein \`a postuler les \'equations qui d\'eterminent le champ de
gravitation engendr\'e par une distribution de mati\`ere:
\be\lb{einstein} R_{\mu\nu} - \frac12Rg_{\mu\nu} = \kappa
T_{\mu\nu}\,. \ee Ces \'equations non lin\'eaires aux d\'eriv\'ees
partielles d'ordre 2 qui (nous allons le voir) sont la
g\'en\'eralisation covariante de la loi non-relativiste
d\'eterminant le potentiel newtonien, permettent de d\'eterminer la
g\'eom\'etrie courbe de l'espace-temps (membre de gauche) \`a partir
de la distribution d'impulsion-\'energie de la mati\`ere (membre de
droite), $\kappa$ \'etant la constante de gravitation d'Einstein.

Pour d\'eterminer la constante $\kappa$, remarquons d'abord que la
contraction des \'equations d'Einstein (\ref{einstein}) conduit \`a
$-R = \kappa T$, ce qui permet de r\'e\'ecrire ces \'equations sous
la forme $$R_{\mu\nu}  = \kappa\left(T_{\mu\nu} -
\frac12Tg_{\mu\nu}\right)\,.$$ Il en r\'esulte, \`a l'approximation
non-relativiste, \be R_{00} \simeq \partial_i\Gamma_{00}^i \simeq
-\frac12\nabla^2g_{00} \simeq \nabla^2U \simeq \frac{\kappa}2T_{00}
= 4\pi GT_{00}\,, \ee o\`u nous avons identifi\'e $T_{00}$ \`a la
densit\'e de masse $\rho$, et utilis\'e la loi de Newton $\nabla^2U
= 4\pi G\rho$ d\'eterminant le potentiel de gravitation engendr\'e
par cette densit\'e. La constante d'Einstein est donc reli\'ee \`a
la constante de Newton par \be \kappa = \frac{8\pi G}{c^4} \ee (o\`u
nous avons r\'etabli la vitesse de la lumi\`ere $c$).

Les \'equations d'Einstein (\ref{einstein}) peuvent \^etre
d\'eriv\'ees en extr\'emisant l'action \be\lb{actrg} S = S_g + S_m
\,, \ee o\`u l'action $S_g$ de la gravitation est l'action
d'Einstein-Hilbert \be S_g = -\frac1{2\kappa}\int R\sqrt{|g|}\,d^4x
\ee ($\sqrt{|g|}\,d^4x$, avec $g =$ det($g_{\mu\nu}$), est la
densit\'e invariante de 4-volume en coordonn\'ees curvilignes), et
$S_m$ est l'action de la mati\`ere.

Pour calculer la variation de l'action d'Einstein-Hilbert lors d'une
variation $\delta g_{\mu\nu}$ du tenseur m\'etrique, d\'ecomposons
\be \delta(R\sqrt{|g|}) = \sqrt{|g|}\left[g^{\mu\nu}\delta
R_{\mu\nu} + R_{\mu\nu}\delta g^{\mu\nu} +
\frac12R\frac{\delta|g|}{|g|}\right]\,.\ee Le premier terme conduit
\`a une divergence qui disparait apr\`es int\'egration par parties,
car \ba \sqrt{|g|}g^{\mu\nu}\delta R_{\mu\nu} &=&
\sqrt{|g|}g^{\mu\nu}\left[D_{\lambda}\delta\Gamma_{\mu\nu}^{\lambda}
- D_{\nu}\delta\Gamma_{\mu\lambda}^{\lambda}\right] \nn\\
&=& D_{\lambda}(\sqrt{|g|}g^{\mu\nu}\delta\Gamma_{\mu\nu}^{\lambda})
- D_{\nu}(\sqrt{|g|}g^{\mu\nu}\delta\Gamma_{\mu\lambda}^{\lambda})
\nn\\ &=&
\partial_{\lambda}(\sqrt{|g|}g^{\mu\nu}\delta\Gamma_{\mu\nu}^{\lambda})
-
\partial_{\nu}(\sqrt{|g|}g^{\mu\nu}\delta\Gamma_{\mu\lambda}^{\lambda})
\approx 0\,, \ea o\`u nous avons utilis\'e $D_{\lambda}g^{\mu\nu} =
0$, et \be D_{\lambda}A^{\lambda} =
\frac1{\sqrt{|g|}}\partial_{\lambda}(\sqrt{|g|}A^{\lambda})\,. \ee
En remarquant que \be \delta g^{\mu\nu} = -
g^{\mu\lambda}g^{\nu\rho}\delta g_{\lambda\rho}\,, \quad
\frac{\delta|g|}{|g|} = g^{\mu\nu}\delta g_{\mu\nu}\,, \ee il reste
\be \delta S_g = \frac1{2\kappa}\int\left[R^{\mu\nu} -
\frac12Rg^{\mu\nu}\right]\delta g_{\mu\nu}\sqrt{|g|}\,d^4x \,. \ee

La variation de l'action de la mati\`ere \be S_m = \int {\cal L}_m
\sqrt{|g|}\,d^4x \ee \'etant, par d\'efinition du tenseur
d'impulsion-\'energie, \be \delta S_m \equiv -\frac12\int T^{\mu\nu}
\delta g_{\mu\nu}\sqrt{|g|}\,d^4x \,, \ee la variation de l'action
totale (\ref{actrg}) conduit bien aux \'equations d'Einstein \be
R_{\mu\nu} - \frac12Rg_{\mu\nu} = \kappa T_{\mu\nu}\,. \ee

\subsubsection{Energie-impulsion et moment angulaire}
En \'electromagn\'etisme, l'int\'egration de l'\'equation de
Maxwell-Gauss \be \nabla\cdot\vec{E} = \rho \ee dans un volume
${\cal V}$ d'espace conduit au th\'eor\`eme de Gauss \be \oint
\vec{E}\cdot d\vec{\Sigma} = Q\,, \ee qui exprime que la charge
\'electrique $Q$ contenue dans ${\cal V}$ est donn\'ee par le flux
du champ \'electrique \`a travers la surface orient\'ee entourant ce
volume. Cette charge est conserv\'ee en vertu de l'\'equation de
continuit\'e $\partial_{\nu}J^{\nu} = 0$.

En relativit\'e g\'en\'erale, la d\'efinition des charges
conserv\'ees $P^{\mu}$ (\'energie pour $\mu=0$ et impulsion pour
$\mu=i$) et $J^{ij}$ (moment angulaire) est compliqu\'ee par la
non-lin\'earit\'e des \'equations d'Einstein. A cause de cette
non-lin\'earit\'e, l'\'energie totale contenue dans un volume ${\cal
V}$ est la somme de l'\'energie de la mati\`ere contenue dans ce
volume et de l'\'energie du champ de gravitation qu'elle engendre.
Cette \'energie est assez d\'elicate \`a calculer, mais les
approches modernes fond\'ees sur la notion d'\'energie quasi-locale
(voir en particulier \cite{chennest} et \cite{booth}) conduisent \`a
des r\'esultats sans ambigu\"{i}t\'e.

Dans le cas de champs de gravitation faibles, on peut recourir \`a
une d\'efinition simplifi\'ee fond\'ee sur la lin\'earisation de la
m\'etrique. Supposons qu'\`a grande distance des sources la
m\'etrique est voisine de la m\'etrique de Minkowski, \be g_{\mu\nu}
= \eta_{\mu\nu} + \gamma_{\mu\nu}\,, \ee avec $|\gamma_{\mu\nu}| \ll
1$. Au premier ordre en $\gamma$, \ba \Gamma_{\mu\nu}^{\lambda}
&\simeq& \frac12\left[\partial_{\mu}\gamma_{\nu}^{\lambda} +
\partial_{\nu}\gamma_{\mu}^{\lambda} -
\eta^{\lambda\rho}\partial_{\rho}\gamma_{\mu\nu}\right]\,, \nn\\
R_{\mu\nu} &\simeq&
\frac12\left[\partial_{\mu}\partial_{\lambda}\gamma_{\nu}^{\lambda}
+ \partial_{\nu}\partial_{\lambda}\gamma_{\mu}^{\lambda} -
\eta^{\lambda\rho}\partial_{\lambda}\partial_{\rho}\gamma_{\mu\nu} -
\partial_{\mu}\partial_{\nu}\gamma\right]\,, \ea o\`u les indices
sont \'elev\'es avec la m\'etrique de Minkowski
($\gamma_{\nu}^{\lambda} \equiv
\eta^{\lambda\rho}\gamma_{\nu\rho}$), et $\gamma \equiv
\gamma_{\lambda}^{\lambda}$.

Les identit\'es de Bianchi lin\'earis\'ees \be
\partial_{\nu}\left({R}^{\mu\nu} - \frac12Rg^{\mu\nu}\right) \simeq
0\,. \ee s'int\`egrent par \be {R}^{\mu\nu} - \frac12Rg^{\mu\nu}
\simeq -
\partial_{\lambda}Q^{\lambda\mu\nu}\,, \ee o\`u $Q^{\lambda\mu\nu}$,
antisym\'etrique en ($\lambda,\mu$), et sym\'etrique en ($\mu,\nu$),
a pour expression \be Q^{\lambda\mu\nu} \equiv
\partial^{[\lambda}\gamma^{\mu]\nu} +
\eta^{[\lambda\nu}(\partial^{\mu]}\gamma -
\partial_{\rho}\gamma^{\mu]\rho}) \ee (o\`u on antisym\'etrise dans
les indices entre crochets, $[\lambda..\mu] = \frac12(\lambda..\mu -
\mu..\lambda)$).

Toujours au premier ordre en $\gamma$, seule la mati\`ere contribue
\`a la quadri-impulsion \be P^{\mu} \simeq \int T^{0\mu}d^3x = -
\frac1{\kappa}\int \partial_{\lambda}Q^{\lambda0\mu}d^3x = -
\frac1{\kappa}\oint Q^{i0\mu}d\Sigma_i\,. \ee En particulier,
l'\'energie totale contenue dans le volume ${\cal V}$ est donn\'ee
par \be P^0 = \frac1{\kappa}\oint (\partial_j\gamma_{ij} -
\partial_i\gamma_{jj})d\Sigma_i\,. \ee De m\^eme, le moment
angulaire est donn\'e par \ba J^{ij} &=& \int (x^iT^{0j} -
x^jT^{0i})d^3x \nn\\ &=& -
\frac1{\kappa}\int(x^i\partial_{\lambda}Q^{\lambda0j} -
x^j\partial_{\lambda}Q^{\lambda0i}) d^3x \nn\\ &=&
-\frac1{\kappa}\oint (x^iQ^{k0j} - x^jQ^{k0i})d\Sigma_k +
\frac2{\kappa}\int Q^{[i0j]}d^3x\,, \ea o\`u le premier terme est le
moment orbital total, et le deuxi\`eme est le moment angulaire
intrins\`eque ou ``spin''. Ce dernier terme peut s'\'ecrire \ba
S^{ij} &=& \frac1{2\kappa}\int (\partial^i\gamma^{0j} -
\partial^j\gamma^{0i})d^3x \nn\\ &=& - \frac1{2\kappa}\oint
(\delta_{ik}\gamma_{0j} - \delta_{jk}\gamma_{0i})d\Sigma_k\,. \ea

\section{Trous noirs stationnaires}
\setcounter{equation}{0}
\subsection{Introduction}
En th\'eorie newtonienne, l'\'energie non-relativiste d'un corps d'\'epreuve
de masse $m$ \`a la distance $r$ d'une masse ponctuelle $M$ est
\be
E = \frac12mv^2 - \frac{GMm}r\,.
\ee
Le corps d'\'epreuve peut donc s'\'echapper du champ gravitationnel de la
masse $M$ si sa vitesse est sup\'erieure \`a la {\em vitesse de lib\'eration}
$v_L = (2GM/r)^{1/2}$. Cette vitesse devient \'egale \`a la vitesse de la
lumi\`ere, $v_L = c$, \`a la distance $r = R_0 = 2GM/c^2$. Il en r\'esulte,
comme montr\'e par Michell (1783) et Laplace\footnote{La traduction de
l'article de Laplace est reproduite en Appendice du livre \cite{hawell}.}
(1796), que la lumi\`ere ne peut pas s'\'echapper d'une \'etoile de rayon
$R < R_0$, qui est donc invisible pour un observateur ext\'erieur (\'etoile
noire).

Cet argument th\'eorique a \'et\'e confirm\'e dans le cadre de la th\'eorie
de la relativit\'e g\'en\'erale. Un bref historique:
\begin{itemize}
\item Schwarzschild (1916) a obtenu la solution statique \`a sym\'etrie sph\'erique des
\'equations d'Einstein, aujourd'hui interpr\'et\'ee comme repr\'esentant un
trou noir.
\item Chandrasekhar (1930) a d\'ecrit la formation de trous noirs par
effondrement gravitationnel.
\item La notion d'horizon \'ev\`enement est due \`a Finkelstein (1958).
\item Le terme de ``trou noir'' semble avoir \'et\'e employ\'e pour la
premi\`ere fois par Wheeler (1967).
\end{itemize}

Du point de vue des valeurs num\'eriques, le calcul fait pour le
Soleil avec $G = 6.67\times10^{-11}$ m$^3$kg$^{-1}$s$^{-2}$ et
$M_{\odot} = 1.99\times 10^{30}$kg donne $R_0 = 2.95\times10^3$m
$\ll R_{\odot} = 6.96\times10^8$m. Notre Soleil n'est donc pas une
\'etoile noire! Dans le cas de motre Galaxie, le ``rayon noir'' vaut
$R_0 = 2.7\times 10^{14}$m $= 0.028$ ann\'ee-lumi\`ere, grand devant
le rayon $R_{\odot}$ du Soeil, mais petit devant le rayon de la
Galaxie $R_G \simeq 4.8\times10^{20}$m $\simeq 50000$
ann\'ees-lumi\`ere. L'observation astrophysique montre qu'il existe
des trous noirs supermassifs au centre des galaxies, dont la masse
est de l'ordre de $10^5$ \`a $10^{10}$ masses solaires.On observe
\'egalement des trous noirs r\'esultant de l'effondrement
gravitationnel d'une \'etoile. Le plus petit r\'ecemment observ\'e
(2008) a une masse de l'ordre de 4 masses solaires.

Les trous noirs dans l'Univers se forment par effondrement gravitationnel et
croissent par accr\'etion de mati\`ere, donc sont en constante \'evolution
dynamique. Cependant, un trou noir suffisamment isol\'e (loin d'autres corps
mat\'eriels et d'autres trous noirs) et suffisamment froid (voir 2.9 plus loin)
approchera au bout d'un certain temps un \'etat quasi-stationnaire, bien
d\'ecrit par une solution exacte stationnaire des \'equations d'Einstein (ou
d'Einstein-Maxwell dans le cas de trous noirs charg\'es \'electriquement).
C'est \`a la g\'eom\'etrie de ces trous noirs stationnaires que nous allons
nous int\'eresser maintenant.

\subsection{Solution de Schwarzschild}
Quelle est la solution des \'equations d'Einstein dans le vide \be
R_{\mu\nu}(x) - \frac12R(x)g_{\mu\nu}(x) = 0 \quad ? \ee La
r\'eponse \`a cette question d\'epend du domaine de d\'efinition de
$x = (t,\vec{x})$, et des conditions aux limites.

Si $x \in R\times R^3$ (tout l'espace-temps), et
$g_{\mu\nu}(t,\vec{x}) {\longrightarrow} \eta_{\mu\nu}$ pour $|x|
\to \infty$ (espace-temps asymptotiquement plat), la seule solution
est Minkowski, $g_{\mu\nu}(x) = \eta_{\mu\nu}$.

Si $x \in R\times R^3$, mais on impose seulement la condition que le champ
de gravitation (tenseur de Riemann) est faible \`a l'infini, on peut avoir
(comme dans le cas des \'equations de Maxwell) des solutions ondulatoires,
d\'ecrivant la propagation d'ondes gravitationnelles.

Plus g\'en\'eralement, se pose la question de savoir quelle est la solution
$g_{\mu\nu}(x)$ des \'equations d'Einstein, non plus dans tout l'espace-temps,
mais \`a l'ext\'erieur de sources  mat\'erielles $T_{\mu\nu}(x)$ localis\'ees
dans un domaine d'espace $\cal D$ compact. A grande distance des sources, on
peut souvent approximer la distribution de mati\`ere par une distribution \`a
sym\'etrie sph\'erique. Dans ce cas on a le

\subsubsection{Th\'eor\`eme de Birkhoff}
Ce th\'eor\`eme (1923) \'enonce que la seule solution \`a sym\'etrie
sph\'erique des \'equations d'Einstein dans le vide est n\'ecessairement
statique,
\be\lb{stat}
ds^2 = -g_{00}(\vec{x})dt^2 + g_{ij}(\vec{x})dx^idx^j,
\ee
et asymptotiquement plate. C'est la solution de Schwarzschild (1916)
\be\lb{S}
ds^2 = -\left(1-\frac{2GM}r\right)dt^2 + \left(1-\frac{2GM}r\right)^{-1}dr^2
+ r^2(d\theta^2 + \sin^2\theta d\varphi^2)
\ee
\`a une transformation de coordonn\'ees pr\`es.

Le potentiel $-g_{00} = 1 - 2GM/r$ correspond bien au potentiel newtonien si
la constante d'int\'egration $M$ est la masse de la source. Cette solution
est valable par exemple \`a l'ext\'erieur d'une \'etoile sph\'erique de masse
$M$, qu'elle soit statique ou en effondrement gravitationnel.

Un corollaire de ce th\'eor\`eme est que, comme dans le cas de
l'\'electromagn\'etisme, les solutions ondulatoires (non statiques) sont
n\'ecessairement \`a sym\'etrie non sph\'erique, donc multipolaires
(dipolaires dans le cas des ondes \'electromagn\'etique, quadrupolaires dans
le cas des ondes gravitationnelles).

La m\'etrique de Schwarzschild (\ref{S}) est {\em singuli\`ere},
d'une part au point $r = 0$ (o\`u $g_{tt} = \infty, g_{ij} = 0$),
d'autre part sur la sph\`ere $r = R_S \equiv 2GM$ (o\`u $g_{tt} = 0,
g_{rr} = \infty$). On sait aujourd'hui que cette singularit\'e de
Schwarzschild est une singularit\'e fictive, qui peut \^etre
\'elimin\'ee par une transformation de coordonn\'ees (voir 2.3 plus
loin). A la surface de l'\'etoile, la solution de Schwarzschild se
raccorde avec la solution int\'erieure, qui est r\'eguli\`ere. Donc,
en tout \'etat de cause, la solution globale pour une \'etoile \`a
sym\'etrie sph\'erique n'a pas de singularit\'e m\'etrique si son
rayon est sup\'erieur au rayon de Schwarzschild $R_S$.

\subsubsection{Mouvement g\'eod\'esique}

Plut\^ot que d'utiliser l'\'equation des g\'eod\'esiques \be
\frac{du^{\lambda}}{d\tau} + \Gamma_{\mu\nu}^{\lambda}u^{\mu}u^{\nu}
= 0\,, \ee il est plus simple de partir de l'action \be\lb{acgeo} I
= \frac12\int g_{\mu\nu}(x)\dot{x}^{\mu}\dot{x}^{\nu}d\lambda \qquad
(\dot{x}^{\mu} \equiv dx^{\mu}/d\lambda)\,, \ee o\`u $\lambda$ est
un param\`etre affine, et de commencer par mettre en \'evidence les
constantes du mouvement, ou int\'egrales premi\`eres. Dans le cas de
la solution de Schwarzschild \be ds^2 = -f(r)\,dt^2 +
f(r)^{-1}\,dr^2 + r^2\,d\Omega^2\,, \ee avec
$$f(r) = 1 - \frac{2GM}r\,, \quad d\Omega^2 = d\theta^2 +
\sin^2\theta d\varphi^2\,,$$ l'invariance par translation dans le
temps (solution statique) entra\^ine la conservation de l'\'energie
\be\lb{E} E = -p_t = f(r)\dot{t} \ee (o\`u $p_{\mu} =
g_{\mu\nu}\dot{x}^{\mu}$ est le moment conjugu\'e \`a la variable
$x^{\mu}$). En inversant cette relation, on obtient l'int\'egrale
premi\`ere \be \dot{t} = \frac{E}{f(r)}\,. \ee De m\^eme,
l'invariance par rotations (sym\'etrie sph\'erique) entra\^ine la
conservation du moment angulaire $\vec{L}$, qui assure que la
g\'eod\'esique est dans un plan perpendiculaire \`a $\vec{L}$. En
choisissant ce plan comme plan $\theta = \pi/2$, il reste la
conservation du moment angulaire azimutal \be L_z = L = p_{\varphi}
= r^2\dot{\varphi}\,, \ee conduisant \`a l'int\'egrale premi\`ere
\be \dot{\varphi} = \frac{L}{r^2}\,. \ee

Enfin, le caract\`ere scalaire de l'action (\ref{acgeo}) conduit \`a
l'int\'egrale premi\`ere
\be\lb{mu2}
g_{\mu\nu}(x)\dot{x}^{\mu}\dot{x}^{\nu} \equiv g^{\mu\nu}p_{\mu}p_{\nu} =
\frac{ds^2}{d\lambda^2} = - \mu^2\,,
\ee
o\`u $\mu^2$ est une constante r\'eelle, qui est positive pour des
g\'eod\'esiques de genre temps, nulle pour des g\'eod\'esiques de genre
lumi\`ere, et n\'egative pour des g\'eod\'esiques de genre espace.
V\'erifions la constance de (\ref{mu2}):
\ba
\frac{d(-\mu^2)}{d\lambda^2} &=& 2g^{\mu\nu}\dot{p}_{\mu}p_{\nu} +
\dot{g}^{\alpha\beta}p_{\alpha}p_{\beta} \nn\\
&=& g^{\mu\nu}\partial_{\mu}g_{\lambda\rho}\dot{x}^{\lambda}\dot{x}^{\rho}
p_{\nu} + \partial_{\mu}g^{\alpha\beta}\dot{x}^{\mu}p_{\alpha}p_{\beta} \nn\\
&=& g^{\mu\nu}\partial_{\mu}g_{\lambda\rho}g^{\lambda\alpha}p_{\alpha}
g^{\rho\beta}p_{\beta}p_{\nu} + \partial_{\mu}g^{\alpha\beta}g^{\mu\nu}p_{\nu}
p_{\alpha}p_{\beta} = 0 \,.
\ea
[Pour passer de la premi\`ere \`a la deuxi\`eme ligne, on a utilis\'e
l'\'equation d'Euler-Lagrange $\dot{p}_{\mu} = (1/2)\partial_{\mu}
g_{\lambda\rho}\dot{x}^{\lambda}\dot{x}^{\rho}$, et pour conclure \`a la
nullit\'e de la troisi\`eme ligne on a remarqu\'e que $\partial_{\mu}
g_{\lambda\rho}g^{\lambda\alpha} = -g_{\lambda\rho}\partial_{\mu}
g^{\lambda\alpha}$.]

Pour la m\'etrique de Schwarzschild, l'int\'egrale premi\`ere
(\ref{mu2}) s'\'ecrit \be -\frac{E^2}{f} + \frac{\dot{r}^2}{f} +
\frac{L^2}{r^2} = - \mu^2\,, \ee d'o\`u \be \dot{r}^2 = E^2 -
f(r)\left(\mu^2 + \frac{L^2}{r^2}\right) = E^2 - V(r)\,. \ee Nous
avons ainsi ramen\'e l'\'etude du mouvement g\'eod\'esique dans la
m\'etrique de Schwarzschild \`a celle du mouvement d'une particule
d'\'epreuve non-relativiste dans le potentiel central effectif
$V(r)$.

Dans le cas de particules suivant une trajectoire radiale ($L=0$,
$\mu^2 > 0$), le potentiel effectif varie comme le potentiel
newtonien. La particule d'\'epreuve en chute libre tombe sur la
singularit\'e de Schwarzschild en un temps propre fini (ou un
param\`etre affine fini pour $\mu^2 = 0$). Mais le temps
coordonn\'ee (mesur\'e par l'observateur \`a l'infini) est,
d'apr\`es (\ref{E}), $t = Er^{\ast}$, o\`u \be\lb{tort} r^{\ast} =
\int f^{-1}dr = r + 2GM\ln\left\vert \frac{r}{2GM}-1 \right\vert \ee
est la coordonn\'ee tortue (``tortoise coordinate''). Pour
l'observateur, la particule d'\'epreuve (telle la tortue d'Achille)
n'atteint la sph\`ere de Schwarzschild qu'au bout d'un temps
coordonn\'ee infini.

Puis, la particule d'\'epreuve continue sa chute sur la singularit\'e centrale
$r=0$ qu'elle atteint en un temps propre fini (pour $\mu^2 > 0$)
\be
\tau = \int_0^{2GM} \frac{r^{1/2}dr}{\sqrt{r(E^2/\mu^2-1)+2GM}}\,,
\ee
ou un param\`etre affine fini (pour $\mu^2=0$)
\be
\lambda = \int_0^{2GM} \frac{dr}E = \frac{2GM}E\,.
\ee

La r\'egion $r < 2GM$ a donc une r\'ealit\'e physique (elle peut
\^etre explor\'ee par des observateurs en chute libre) qui \'echappe
aux observateurs ext\'erieurs: la sph\`ere de Schwarzschild $r =
2GM$ est un {\em horizon}. Cet horizon est {\em r\'egulier}: on
montre que le scalaire de Kretschmann
$R^{\mu\nu\rho\sigma}R_{\mu\nu\rho\sigma}$ y reste fini. Par contre,
les g\'eod\'esiques se terminent \`a la singularit\'e centrale $r =
0$, o\`u le scalaire de Kretschmann diverge.

\subsection{Coordonn\'ees de Kruskal}

Pour $r > R_S$, la signature de la m\'etrique (\ref{S}) est $(- + + +)$.
Pour $r < R_S$, la signature de cette m\'etrique est $(+ - + +)$, ce
qui signifie que la coordonn\'ee radiale $r$ est devenue de genre temps: le
c\^one de lumi\`ere a bascul\'e!

La ligne d'univers $r = R_S$ (\`a $\theta$ et $\varphi$ fix\'es) est
du genre lumi\`ere: \be ds^2 = - f(R_S)dt^2 = 0\,. \ee Cette
observation sugg\`ere que la transformation \`a des coordonn\'ees du
type ``c\^one de lumi\`ere'' pourrait r\'egulariser la solution de
Schwarzschild au voisinage de $r=R_S$.

Dans le cas de la m\'etrique de Minkowski (en coordonn\'ees d'espace
sph\'eriques)
\be
ds^2 = -dt^2 + dr^2 + r^2\,d\Omega^2\,,
\ee
la transformation aux coordonn\'ees du c\^one de lumi\`ere
\ba
u &=& t-r\,, \nn\\
v &=& t+r\,,
\ea
conduit \`a la m\'etrique
\be
ds^2 = -du\,dv + \frac{(u-v)^2}4\,d\Omega^2\,.
\ee

Dans le cas de la  m\'etrique de Schwarzschild, \ba
ds^2 &=& -f\,dt^2 + f^{-1}\,dr^2 + r^2\,d\Omega^2 \nn\\
&=& f(-dt^2 + dr^{\ast2}) + r^2\,d\Omega^2 \ea ($r^{\ast}$ \'etant
la coordonn\'ee tortue (\ref{tort})), la transformation de
coordonn\'ees \ba
u &=& t-r^{\ast}\,, \nn\\
v &=& t+r^{\ast}\,,
\ea
conduit \`a
\be\lb{Suv}
ds^2 = -f\,du\,dv + r^2\,d\Omega^2\,,
\ee
o\`u $f$ et $r^2$ sont exprim\'es en fonction de $u$ et $v$.

Effectuons maintenant dans le plan $(u,\,v)$ la transformation
conforme \be(u,\,v) \longrightarrow (U = -F(-u),\,V = F(v))\,. \ee
La m\'etrique (\ref{Suv}) devient \be\lb{SUV} ds^2 =
-\frac{f}{F'(-u)F'(v)}\,dU\,dV + r^2\,d\Omega^2\,. \ee Le choix \be
F(v) = 4GM\e^{\ds v/4GM} \ee conduit \`a \ba \frac{f}{F'(-u)F'(v)}
&=& \e^{\ds(u-v)/4GM}\left(1-\frac{2GM}r\right) =
\e^{\ds-r^{\ast}/2GM}\left(1-\frac{2GM}r\right) \nn\\
&=& \e^{\ds-r/2GM}\frac{2GM}{|r-2GM|}\frac{r-2GM}r = \frac{2GM}r
\e^{\ds-r/2GM}\,,
\ea
o\`u nous avons utilis\'e (\ref{tort}), et suppos\'e $r > 2GM$. Dans une
derni\`ere \'etape, repassons des coordonn\'ees du c\^one de lumi\`ere $U,\,V$
\`a des coordonn\'ees 2-minkowskiennes $(T,\,R)$,
\ba
U &=& T-R\,, \nn\\
V &=& T+R\,, \ea pour aboutir \`a la forme de Kruskal de la solution
de Schwarzschild \be\lb{krusk} ds^2 =
\frac{2GM}r\e^{\ds-r/2GM}(-dT^2+dR^2) + r^2\,d\Omega^2\,, \ee qui
est r\'eguli\`ere en $r = 2GM$! Nous avons donc, comme annonc\'e,
\'elimin\'e la singularit\'e de Schwarzschild par une transformation
de coordonn\'ees.

Quand $r$ varie de $2GM$ \`a l'infini, la coordonn\'ee tortue $r^{\ast}$ varie,
d'apr\`es (\ref{tort}), de $-\infty$ \`a $+\infty$. Donc, tandis que les
coordonn\'ees de genre lumi\`ere $u$ et $v$ varient chacune sur la droite
r\'eelle, le domaine des coordonn\'ees $U$ et $V$ est
$$ U \in R_-\,, \qquad V \in R_+ \,. $$
C'est le domaine $I$ du diagramme de la figure 1.
\begin{figure}
\centering
\includegraphics[width=5cm,angle=0]{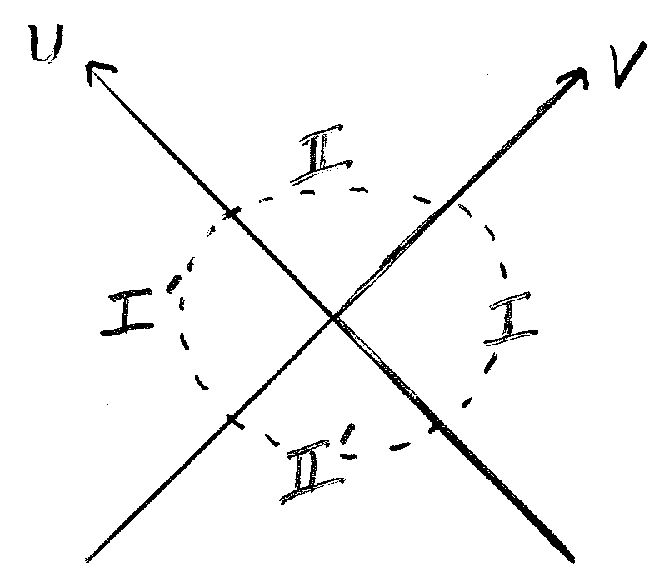}
\caption{Domaines de Kruskal} \label{fig1}
\end{figure}

Les fronti\`eres de ce domaine
$$ \begin{array}{cccccc} V = 0 & (T=-R) &\quad & v = t+r^{\ast} & \to &
-\infty \\ U = 0 & (T=+R) &\quad & u = t-r^{\ast} & \to & +\infty \end{array}$$
correspondent toutes les deux \`a l'horizon $r^{\ast} \to -\infty$ (\`a $t$
fix\'e). La m\'etrique de Kruskal (\ref{krusk}) peut \^etre prolong\'ee \`a
travers ces fronti\`eres, mais pas la transformation de coordonn\'ees de
Kruskal qui nous a permis de l'obtenir!

En prolongeant la m\'etrique de Kruskal \`a travers la branche $U=0$
($V>0$) de l'horizon, on passe dans la r\'egion $II$ ($U>0$, $V>0$).
Cette r\'egion est \`a l'int\'erieur de l'horizon, donc doit
correspondre au domaine $r<2GM$ de la m\'etrique de Schwarzschild.
Son autre fronti\`ere $V=0$ ($U>0$) \'etant \`a nouveau une branche
de l'horizon, on peut encore prolonger la m\'etrique de Kruskal \`a
travers cette fronti\`ere pour passer dans une r\'egion $I'$ ($U>0$,
$V<0$) ext\'erieure \`a l'horizon ($r > 2GM$). De proche en proche,
on montre ainsi que la m\'etrique de Kruskal couvre les quatre
secteurs $I$, $I'$ (ext\'erieur, $r > 2GM$) et $II$, $II'$
(int\'erieur, $r < 2GM$).

La transformation de coordonn\'ees entre la m\'etrique de
Schwarzschild et la m\'etrique de Kruskal peut \^etre \'ecrite de
fa\c{c}on compacte (pour les quatre secteurs): \be\lb{epset} U =
-\epsilon F(-u)\,, \quad V = \eta F(v) \qquad (\epsilon, \eta = \pm
1)\,. \ee Il en r\'esulte \ba
fdudv &=& \epsilon\,\eta\,\e^{\ds-r^{\ast}/2GM}\left(1-\frac{2GM}r\right)dUdV\nn\\
&=& \epsilon\,\eta\,\mbox{\rm sign}(r-2GM)\frac{2GM}r\e^{\ds-r/2GM}dUdV\,.
\ea
Donc la r\'egion ext\'erieure $r>2GM$ correspond \`a $\eta = \epsilon$, et
la r\'egion int\'erieure $r<2GM$ correspond \`a $\eta = -\epsilon$, comme
montr\'e sur la figure 2.
\begin{figure}
\centering
\includegraphics[width=5cm,angle=0]{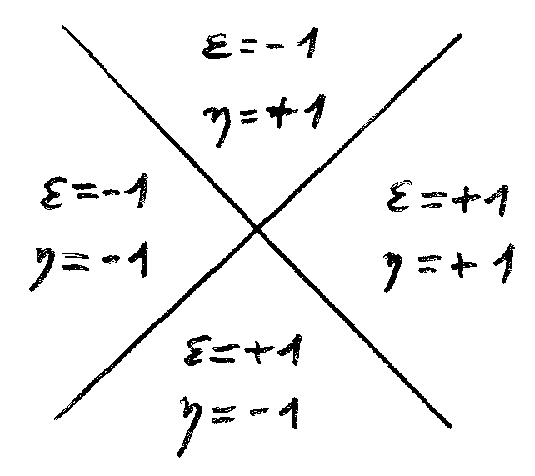}
\caption{Transformations de Kruskal (\ref{epset}) dans les 4
domaines de Kruskal} \label{fig2}
\end{figure}

Les coordonn\'ees de Kruskal r\'ealisent l'{\em extension analytique
maximale} de la solution de Schwarzschild. La carte $S$
(Schwarzschild) est d\'efinie en dehors (\`a l'ext\'erieur ou \`a
l'int\'erieur) de l'horizon, pour chacun des secteurs $I$, $I'$,
$II$, $II'$. Elle est {\em bonne}{ pour l'observateur ext\'erieur
(\`a l'infini), mais mauvaise pour l'observateur en chute libre. La
carte $K$ (Kruskal) d\'efinie partout est {\em bonne} pour
l'observateur en chute libre, mais mauvaise pour l'observateur
ext\'erieur. La g\'eom\'etrie de cet espace-temps peut \^etre mieux
visualis\'ee \`a l'aide de diagrammes de Penrose.

\subsection{Diagrammes de Penrose}

Revenons \`a la m\'etrique de Minkowski en coordonn\'ees du c\^one de
lumi\`ere,
\be\lb{min1}
ds^2 = -du\,dv + \frac{(u-v)^2}4\,d\Omega^2 \qquad (u = t-r\,, \; v = t+r)\,,
\ee
et faisons la transformation conforme \`a deux dimensions
\be
u = \ell\tan{p}\,, \quad v = \ell\tan{q}\,, \quad \mbox{\rm avec} -\pi/2 <
(p, \; q) < \pi/2\,,
\ee
o\`u $\ell$ est une constante. La condition $r = (v-u)/2 \ge 0$ entra\^ine
$q \ge p$. La m\'etrique de Minkowski (\ref{min1}) devient
\be ds^2 = \frac{\ell^2}{4\cos^2p\cos^2q}\bigg[-4dp\,dq + \sin^2(p-q)\,
d\Omega^2\bigg]\,.
\ee
Puis la nouvelle transformation
\be
p = \frac{\psi-\xi}2\,, \; q = \frac{\psi+\xi}2\,, \quad (0 \le \xi \le \pi\,,
\; \xi - \pi \le \psi \le \pi - \xi)\,,
\ee
conduit \`a la m\'etrique
\be\lb{min3}
ds^2 = \frac{\ell^2}{(\cos\psi+\cos\xi)^2}\bigg[-d\psi^2+d\xi^2+\sin^2\xi\,
d\Omega^2\bigg]\,.
\ee
Dans le cas o\`u $\psi \in R$, la m\'etrique entre crochets est celle de
l'univers statique d'Einstein (le produit direct de l'axe des temps par la
sph\`ere \`a trois dimensions $S^3$). La forme de la m\'etrique (\ref{min3})
montre donc que l'espace-temps de Minkowski est conforme \`a l'espace-temps
``cylindrique'' d'Einstein (avec temps $\psi$ compactifi\'e) avec le facteur
conforme $\ell^2/(\cos\psi+\cos\xi)^2$.

Le domaine de variation des angles $\psi$ et $\xi$ est repr\'esent\'e sur le
diagramme de Penrose de la Figure 3. Chaque point de ce diagramme (sauf la
ligne de genre temps $\xi = 0$, qui correspond \`a $u = v$, donc \`a l'origine
des coordonn\'ees d'espace $r=0$) est une 2-sph\`ere. Les autres fronti\`eres
Scri$_+$ ($\psi = \pi - \xi$, ou $v = +\infty$) et Scri$_-$ ($\psi = \xi -
\pi$, ou $u = -\infty$), ainsi que leur point d'intersection $i_0$
correspondent \`a l'infini de genre espace $r \to \infty$, tandis que leurs
points d'intersection $i_-$ et $i_+$ avec la droite $\xi = 0$ correspondent
respectivement \`a l'infini pass\'e $t \to -\infty$ et \`a l'infini futur
$t \to +\infty$.
\begin{figure}
\centering
\includegraphics[width=5cm,angle=0]{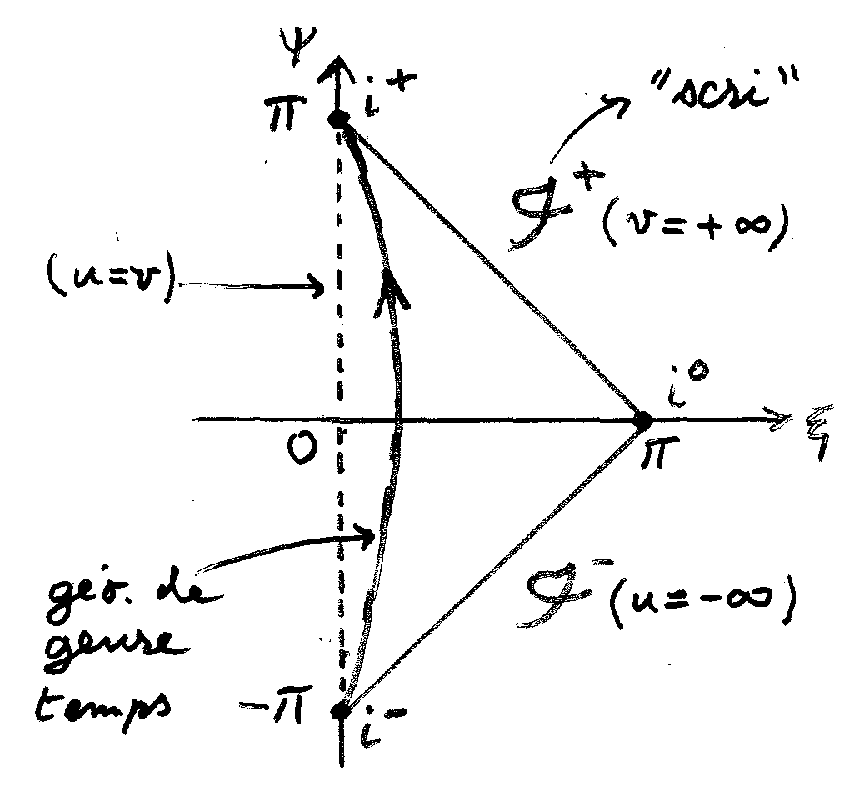}
\caption{Diagramme de Penrose de Minkowski} \label{fig3}
\end{figure}

Les transformations conformes \`a deux dimensions qui nous ont
permis de passer de (\ref{min1}) \`a (\ref{min3}) conservent le
genre des g\'eod\'esiques radiales (celles passant par l'origine
$\xi = 0$). On montre que:
\begin{itemize}
\item  les g\'eod\'esiques quelconques de genre temps commencent \`a l'infini
pass\'e $i_-$ et finissent \`a l'infini futur $i_+$;
\item les g\'eod\'esiques quelconques de genre lumi\`ere commencent \`a
Scri$_-$ et finissent \`a Scri$_+$;
\item les g\'eod\'esiques quelconques de genre espace commencent et finissent
\`a l'infini d'espace $i_0$.
\end{itemize}

Le diagramme de Penrose pour l'extension analytique maximale de la
solution de Schwarzschild est construit de fa\c{c}on analogue.
Partons de la m\'etrique de Kruskal \be ds^2 =
\frac{2GM}r\e^{\ds-r/2GM}(-dU\,dV) + r^2\,d\Omega^2\,, \ee et
effectuons la transformation \be U =
2GM\tan\left(\frac{\psi-\xi}2\right)\,, \quad V = 2GM\tan\left(
\frac{\psi+\xi}2\right)\,. \ee La coordonn\'ee tortue $r^{\ast}$
variant de $-\infty$ \`a $+\infty$, le domaine $I$ de Kruskal est
analogue, non pas \`a celui de l'espace-temps de Minkowski \`a
quatre dimensions, mais \`a celui de l'espace-temps de Minkowski \`a
deux dimensions (dont la coordonn\'ee d'espace varie entre $-\infty$
et $+\infty$). L'horizon est donc repr\'esent\'e par deux lignes de
genre lumi\`ere sym\'etriques de Scri$_-$ et Scri$_+$, correspondant
\`a $V=0$ et $U=0$. Puis les r\'egions int\'erieures $II$ et $II'$
sont born\'ees par la singularit\'e $r=0$ (de genre espace), qui
correspond aussi, d'apr\`es (\ref{tort}) \`a $r^{\ast}=0$, donc \`a
$$ u = v \; \leftrightarrow \; UV = 1 \; \leftrightarrow \; \psi =
\pm\pi/2\,.$$
\begin{figure}
\centering
\includegraphics[width=8cm,angle=0]{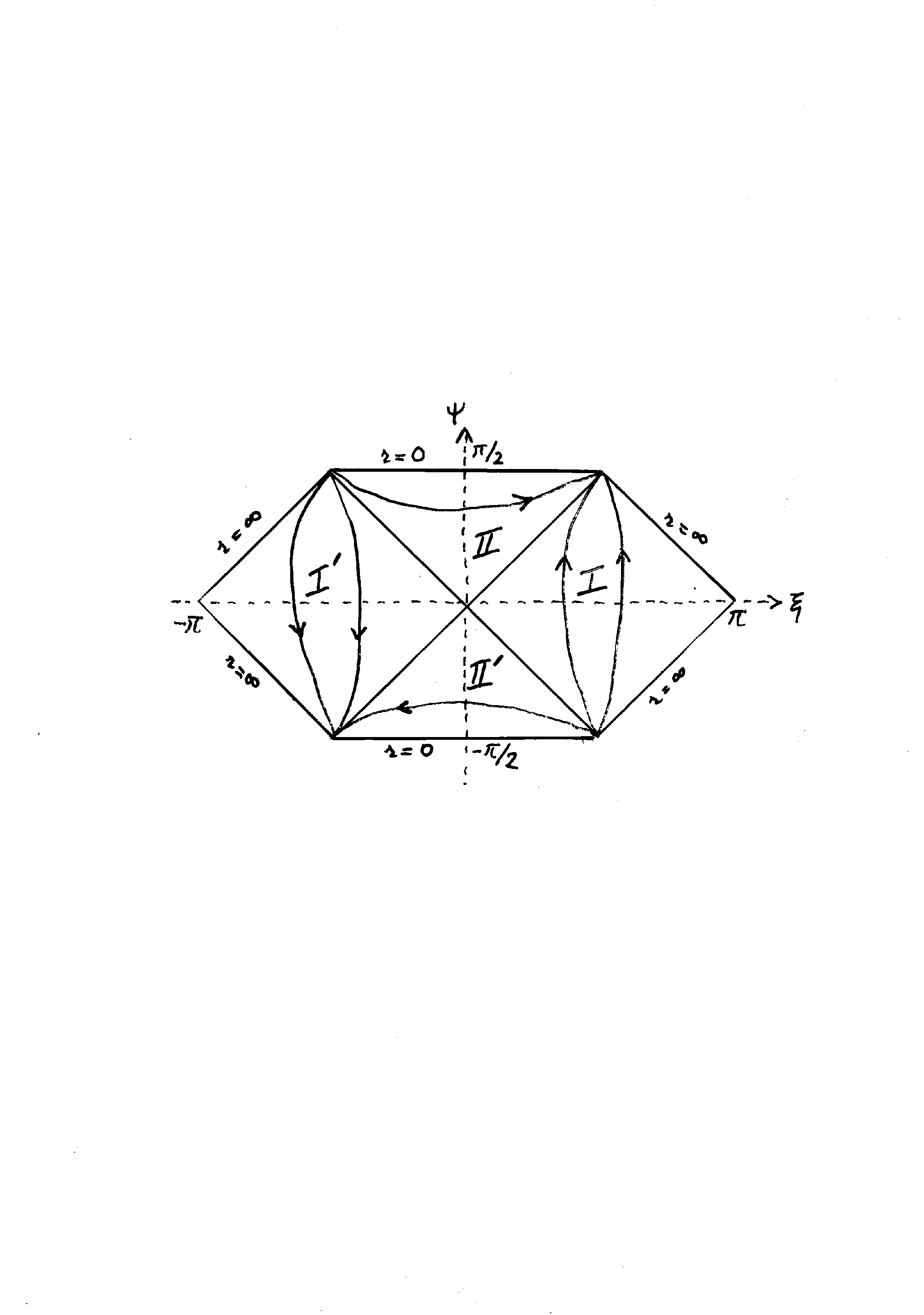}
\caption{Diagramme de Penrose de Schwarzschild} \label{fig4}
\end{figure}
Enfin, le diagramme de Penrose de l'espace-temps de Schwarzschild,
repr\'esent\'e sur la Figure 4, est compl\'et\'e par la r\'egion
ext\'erieure $I'$ sym\'etrique de $I$. Les lignes claires sur ce
diagramme correspondent \`a des sections $r =$ constante,
orient\'ees suivant les temps $t$ croissants.

La Figure 5 repr\'esente deux g\'eod\'esiques de genre temps (lignes
pleines orient\'ees), correspondant aux lignes d'univers
d'observateurs en chute libre \'emettant radialement des signaux
lumineux (lignes ondul\'ees orient\'ees \`a $45^o$). Le premier
observateur est en orbite ($r=$ constante) autour du trou noir, et
ses signaux atteignent toujours l'observateur ext\'erieur, \`a
l'infini Scri$_+$. Le deuxi\`eme observateur n'a pas un moment
angulaire suffisant pour \'eviter de traverser l'horizon. Apr\`es la
travers\'ee de l'horizon, sa chute sur la singularit\'e centrale est
in\'evitable, les seules g\'eod\'esiques reliant la r\'egion $I$ \`a
la r\'egion $I'$ \'etant de genre espace. Ses derniers signaux
\'emis juste avant la travers\'ee de l'horizon arrivent \`a Scri$_+$
(ou \`a l'observateur en orbite) au bout d'un temps infini, tandis
que les signaux \'emis apr\`es la travers\'ee de l'horizon
aboutissent \`a la singularit\'e centrale.
\begin{figure}
\centering
\includegraphics[width=6cm,angle=0]{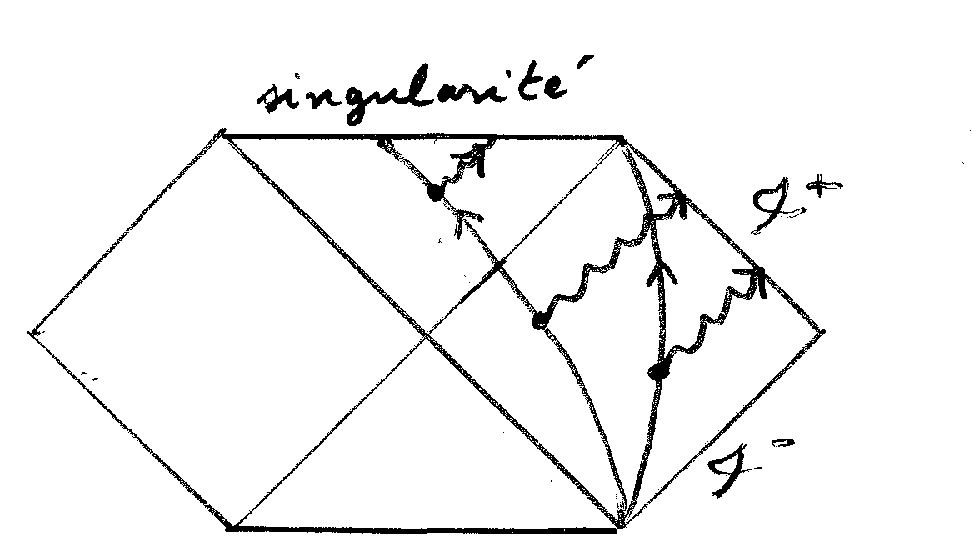}
\caption{G\'eod\'esiques dans l'espace-temps de Schwarzschild}
\label{fig5}
\end{figure}

Cet exemple illustre la {\em D\'efinition} (simplifi\'ee):

\bigskip
{\em Un trou noir est une r\'egion de l'espace-temps d'o\`u les particules et
les photons ne peuvent pas se propager jusqu'\`a l'observateur ext\'erieur
Scri$_+$.}

\subsection{Solution de Reissner-Nordstr\"{o}m}
La th\'eorie de l'\'electromagn\'etisme coupl\'e \`a la gravitation est
d\'efinie par l'action d'Einstein-Maxwell (\'ecrite en unit\'es
gravitationnelles, $G=1$, pour la gravitation et en unit\'es de Gausss pour
l'\'electromagn\'etisme)
\be\lb{acEM}
S = \frac1{16\pi}\int d^4x\sqrt{|g|}\bigg[-R + g^{\mu\rho}g^{\nu\sigma}
F_{\mu\nu}F_{\rho\sigma}\bigg]\,,
\ee
o\`u
\be
F_{\mu\nu} = D_{\mu}A_{\nu} - D_{\nu}A_{\mu} = \partial_{\mu}A_{\nu}
- \partial_{\nu}A_{\mu}
\ee
est le tenseur champ \'electromagn\'etique. Le tenseur d'\'energie-impulsion
\'electromagn\'etique
\be
T_{\mu\nu} = \frac2{\sqrt{|g}}\,\frac{\delta S_m}{\delta g^{\mu\nu}} =
\frac1{4\pi}\bigg[g^{\rho\sigma}F_{\mu\rho}F_{\nu\sigma} - \frac14
g_{\mu\nu}F^{\alpha\beta}F_{\alpha\beta}\bigg]
\ee
est de trace nulle,
\be
T \equiv g^{\mu\nu}T_{\mu\nu} = 0
\ee
(pour $D=4$ seulement). Les \'equations de Maxwell et d'Einstein d\'erivant de
l'action (\ref{acEM}) sont
\ba
D_{\nu}F^{\mu\nu} &=& \frac1{\sqrt{|g|}}\partial_{\nu}\bigg(\sqrt{|g|}
F^{\mu\nu}\bigg) = 0\,, \lb{max}\\
R_{\mu}^{\nu} &=& 2\bigg(F_{\mu\rho}F^{\nu\rho} - \frac14\delta_{\mu}^{\nu}
F_{\rho\sigma}F^{\rho\sigma}\bigg)\lb{einM}\,.
\ea

Pour chercher les solutions \'electrostatiques \`a sym\'etrie
sph\'erique de ce syst\`eme d'\'equations, nous pouvons faire l'{\em
ansatz} \be\lb{an} ds^2 = -f(r) dt^2 + f^{-1}(r) dr^2 +
r^2d\Omega^2\,, \quad A_{\mu} = \delta_{\mu}^0A_0(r)\,, \ee o\`u
$f(r)$ et $A_0(r)$ sont deux fonctions inconnues d'une seule
variable. L'\'el\'ement de volume dans la m\'etrique (\ref{an}) est
$\sqrt{|g|} = r^2\sin\theta$, donc l'\'equation de Maxwell
(\ref{max}) avec $\mu = 0$ s'int\`egre imm\'ediatement par
\be\lb{coulomb} F^{01} = \frac{e}{r^2}\,, \ee o\`u la constante
d'int\'egration $e$ peut \^etre identifi\'ee \`a la charge
\'electrique de la source, comme le montre le calcul du flux \be
\oint F^{0i}\sqrt{|g|}d\Sigma_i = \int\frac{e}{r^2}r^2\sin\theta
d\theta d\varphi = 4\pi e\,. \ee En reportant (\ref{coulomb}) dans
le tenseur d'\'energie-impulsion \'electromagn\'etique, on ram\`ene
les \'equations d'Einstein \`a \be R_0^0 = R_1^1 = -R_2^2 = -R_3^3 =
- \frac{e^2}{r^4}\,, \ee \'equations qui s'int\`egrent par
\be\lb{RN} ds^2 = -\bigg(1-\frac{2m}r + \frac{e^2}{r^2}\bigg)dt^2 +
\bigg(1-\frac{2m}r + \frac{e^2}{r^2}\bigg)^{-1}dr^2 +
r^2d\Omega^2\,, \ee o\`u $m$ est une deuxi\`eme constante
d'int\'egration qui peut, comme dans le cas de la solution de
Schwarzschild, \^etre identifi\'ee \`a la masse de la source (le
potentiel newtonien \`a grande distance est $1 - 2m/r + O(r^{-2})$).
C'est la solution obtenue par Reissner (1916) et Nordstr\"{o}m
(1918).

Ouvrons ici une parenth\`ese pour signaler que cette solution
\'electrostatique a son pendant magn\'etostatique. On peut d\'efinir
le tenseur dual du champ \'electromagn\'etique \be F^{\ast\mu\nu}
\equiv \frac1{2\sqrt{|g|}}\epsilon^{\mu\nu\rho\sigma}
F_{\rho\sigma}\,. \ee Ce champ dual, tel que \be
\partial_{\nu}\bigg(\sqrt{|g|}F^{\ast\mu\nu}\bigg) = 0\,, \quad
T_{\mu\nu}(F^{\ast}) = T_{\mu\nu}(F)\,, \ee v\'erifie les m\^emes
\'equations d'Einstein-Maxwell que le champ \'electrostatique,
conduisant \`a la solution \`a sym\'etrie sph\'erique \be F^{\ast01}
= \frac{e}{r^2} \quad \longleftrightarrow \quad F_{23} =
e\sin\theta\,, \ee avec la m\^eme m\'etrique de
Reissner-Nordstr\"{o}m (\ref{RN}), engendr\'ee par un monop\^ole
magn\'etique $e$ de masse $m$.

Si \underline{$e^2 < m^2$}, la fonction $f(r)$ dans (\ref{RN}) a
deux z\'eros en \be r_{\pm} = m \pm \sqrt{m^2-e^2}\,. \ee Donc la
m\'etrique de Reissner-Nordstr\"{o}m a deux horizons, $r=r_+$
l'horizon \'ev\`enement, et $r=r_-$ l'horizon int\'erieur, ainsi
qu'une singularit\'e centrale en $r = 0 < r_- < r_+$.

Si \underline{$e^2 = m^2$} (solution de Reissner-Nordstr\"{o}m
extr\^eme), la m\'etrique de Reissner-Nordstr\"{o}m a un horizon
double, sans changement de signature.

Enfin, si \underline{$e^2 > m^2$}, la m\'etrique de
Reissner-Nordstr\"{o}m n'a pas d'horizon. On dit dans ce cas que
$r=0$ est une {\em singularit\'e nue}.

Du point de vue num\'erique, la charge \'electrique \'el\'ementaire
est $e = 1.38\times10^{-34}$cm, extr\^emement petite par rapport \`a
la masse du Soleil $m = M_{\odot} = 1.48\times 10^5$cm. Donc, \`a
l'\'echelle astrophysique, d'hypoth\'etiques trous noirs charg\'es
auraient deux horizons. A l'inverse, la masse d'une particule
\'el\'ementaire charg\'ee telle que le proton est $m_p =
1.24\times10^{-52}$cm, tr\`es petite devant sa charge \'electrique,
ce qui justifie qu'on n\'eglige sa self-interaction
gravitationnelle.
\begin{figure}
\centering
\includegraphics[width=10cm,angle=0]{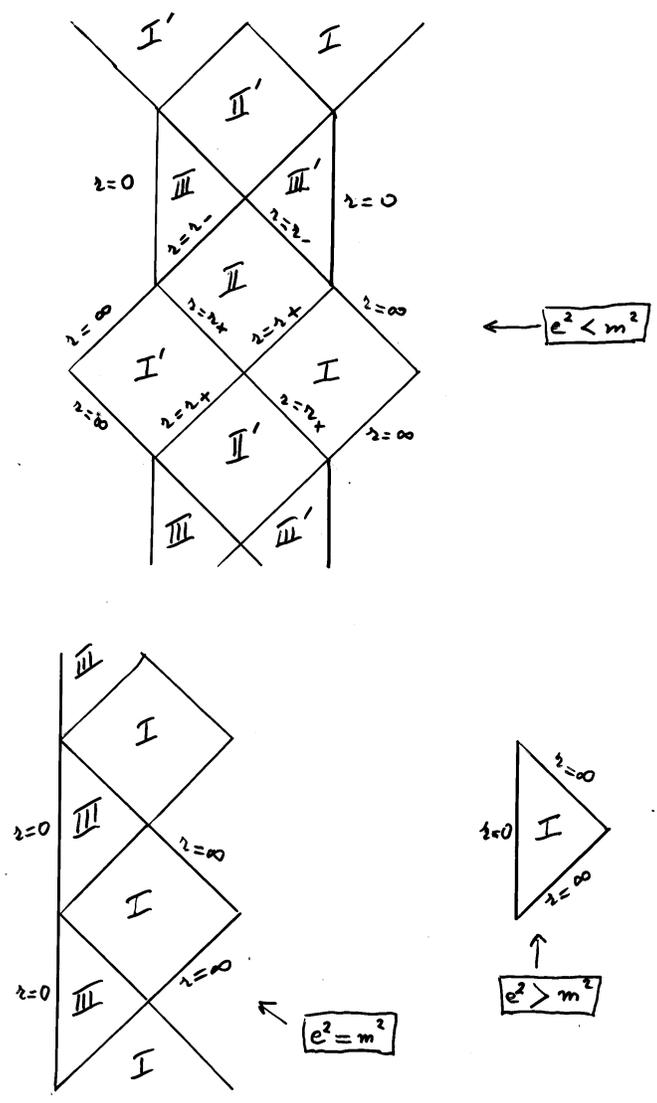}
\caption{Diagrammes de Penrose de Reissner-Nordstr\"{o}m}
\label{fig6}
\end{figure}

La m\'etrique de Reissner-Nordstr\"{o}m peut \^etre prolong\'ee \`a
travers ses deux horizons (dans le cas $e^2<m^2$), ou son horizon
double (dans le cas $e^2=m^2$), par des m\'ethodes analogues \`a
celles employ\'ees pour la m\'etrique de Schwarzschild, et les
diagrammes de Penrose de ses extensions analytiques maximales
peuvent \^etre construits de la m\^eme fa\c{c}on. Ces diagrammes
sont donn\'es dans la Figure 6. Dans le cas $e^2<m^2$, la r\'egion
ext\'erieure $I$ de signature $(-+++)$ se prolonge \`a travers
l'horizon \'ev\`enement $r=r_+$ par une r\'egion int\'erieure $II$
de signature $(+-++)$, qui elle-m\^eme se prolonge \`a travers
l'horizon int\'erieur $r=r_-$ par une deuxi\`eme r\'egion
int\'erieure $III$ de signature $(-+++)$. La singularit\'e centrale
$r=0$ qui limite la r\'egion $III$ est donc (contrairement au cas de
Schwarzschild) de genre temps. Il en r\'esulte que le prolongement
analytique maximal de la m\'etrique de Reissner-Nordstr\"{o}m \`a
travers les diff\'erentes branches de ses horizons conduit \`a une
cha\^{\i}ne infinie de r\'egions $I$, $II$ et $III$ (les r\'egions
$II$ \'etant absentes dans le cas $e^2=m^2$).

\subsection{Multi-trous noirs}

Dans le cas non-relativiste, la force gravitationnelle (de Newton) entre deux
masses
\be
F_g = -\frac{GMM'}{r^2}
\ee
est analogue, mais de signe oppos\'e, \`a la force \'electrostatique
(de Coulomb) entre deux charges
\be
F_e = \frac{QQ'}{4\pi\epsilon_0r^2}\,.
\ee
Les particules avec $Q^2/4\pi\epsilon_0 = GM^2$ sont donc en \'equilibre
statique entre force newtonienne attractive et force coulombienne r\'epulsive
($F = F_g + F_e = 0$) quelle que soit leur position relative.

La relation analogue en relativit\'e g\'en\'erale, $e^2 = m^2$,
caract\'erise la solution de Reissner-Nordstr\"{o}m extr\^eme
\be\lb{RNE} ds^2 = -\bigg(\frac{r-m}r\bigg)^2 +
\bigg(\frac{r}{r-m}\bigg)^2 \bigg[dr^2 + (r-m)^2d\Omega^2\bigg]\,.
\ee En effectuant le changement de coordonn\'ees $r \to r+m$, la
solution de RN extr\^eme prend la forme \ba ds^2 &=&
-\bigg(1+\frac{m}r\bigg)^{-2}dt^2 + \bigg(1+\frac{m}r\bigg)^{2}
d\vec{x}^2\,, \\
A_0 &=& \frac{m}{r+m}\,,
\ea
o\`u (contrairement aux apparences) le ``point'' $r=0$ est en fait une sph\`ere
(l'horizon double) d'aire $4\pi m$.

Sous cette forme, la solution de Reissner-Nordstr\"{o}m extr\^eme
admet une g\'en\'eralisation multi-centres, donn\'ee
ind\'ependemment par Majumdar (1947) et Papapetrou (1947),
\ba\lb{MP}
ds^2 &=& -(1+\sigma)^{-2}dt^2 + (1+\sigma)^{2}d\vec{x}^2\,, \\
A_0 &=& \frac{\sigma}{1+\sigma}\,,
\ea
les \'equations de Maxwell et d'Einstein \'etant simultan\'ement satisfaites
si le potentiel $\sigma(\vec{x})$ est harmonique,
\be
\nabla^2\sigma = 0\,.
\ee
Le laplacien figurant dans cette \'equation est le laplacien ordinaire sur
l'espace plat de m\'etrique ``r\'eduite'' $d\vec{x}^2$ apparaissant dans
(\ref{MP}). Cet op\'erateur \'etant lin\'eaire, on peut superposer un nombre
quelconque de monop\^oles pour obtenir une solution multi-centres (sans
sym\'etrie spatiale)
\be
\sigma = \Sigma_i\frac{m_i}{|\vec{x}-\vec{x_i}|}
\ee
($m_i > 0$, $\vec{x_i}$ quelconques) \`a l'ext\'erieur des horizons $\vec{x} =
\vec{x_i}$, la m\'etrique (\ref{MP}) obtenue pouvant \^etre prolong\'ee \`a
la Kruskal \`a l'int\'erieur de chaque horizon. On peut \'egalement, en
prenant la limite o\`u plusieurs ``centres'' $\vec{x} = \vec{x_i}$
co\"{\i}ncident, ajouter \`a cette superposition des dip\^oles, quadrup\^oles,
etc.

Enfin, la solution de Majumdar-Papapetrou, qui est asymptotiquement
plate, admet la g\'en\'eralisation non-asymptotiquement plate (NAP),
obtenue en effectuant la translation $\sigma \to \sigma-1$ (qui est
encore solution de $\nabla^2\sigma = 0$): \be ds^2 =
-\sigma^{-2}dt^2 + \sigma^2d\vec{x}^2\,, \quad A_0 = -\frac1{\sigma}
\qquad \bigg(\sigma =
\Sigma_i\frac{m_i}{|\vec{x}-\vec{x_i}|}\bigg)\,. \ee

\subsection{Solution de Bertoti-Robinson}
La solution de Majumdar-Papapetrou NAP \`a un centre, d\'erivant du potentiel
harmonique $\sigma = -1/er$, est
\ba
A_{\mu}dx^{\mu} &=& er\,dt\,, \\
ds^2 &=& -e^2r^2\,dt^2 + e^{-2}\frac{dr^2}{r^2} + e^{-2}d\Omega^2\,. \lb{BR0}
\ea
Cette solution appartient \`a une famille \`a un param\`etre de solutions
NAP \`a sym\'etrie sph\'erique, obtenue \`a partir de l'ansatz
\be
ds^2 = -f(r)dt^2 + f(r)^{-1}dr^2 + \ell^2d\Omega^2
\ee
($\ell$ constante). L'\'el\'ement de volume est maintenant $\sqrt{|g|} =
\ell^2\sin\theta$, donc la solution des \'equations de Maxwell est
\be
F^{01} = e
\ee
($e$ constante d'int\'egration).
En reportant dans les \'equations d'Einstein, on obtient le syst\`eme
\be
R_0^0 = R_1^1 = -R_2^2 = -R_3^3 = - e^2\,,
\ee
dont la solution \cite{br} est
\be\lb{BR}
ds^2 = -e^2(r^2-b^2)dt^2 + e^{-2}\frac{dr^2}{r^2-b^2} + e^{-2}(d\theta^2
+ \sin^2\theta d\varphi^2)\,,
\ee
o\`u $e = \ell^{-1}$, et $b^2$ est une constante d'int\'egration r\'eelle.

L'espace-temps de m\'etrique (\ref{BR}) (o\`u nous pouvons choisir l'unit\'e
de longueur $e^{-1} = 1$) est le produit $AdS_2 \times S^2$
de deux espaces \`a courbure constante, \begin{itemize}
\item une sph\`ere $S^2$, qui est la surface $x^2 + y^2 + z^2 = 1$ dans
l'espace euclidien $ds^2 = dx^2 + dy^2 + dz^2$, et
\item un espace d'anti-de Sitter $AdS_2$, qui est la surface $x^2 + y^2
- z^2 = 1$ dans l'espace pseudo-euclidien $ds^2 = - dx^2 - dy^2 + dz^2$.
\end{itemize}
L'espace d'anti-de Sitter s'obtient donc formellement \`a partir de la
sph\`ere par la continuation analytique $z \to iz$, $ds^2 \to -ds^2$. En
partant de la param\'etrisation de la
sph\`ere en coordonn\'ees angulaires ($\theta$, $\varphi$), cette continuation
(qui implique la continuation $\theta \to \pi/2 - i\alpha$) conduit \`a la
param\'etrisation de $AdS_2$ en coordonn\'ees hyperboliques:
\be
S^2 \left\vert\begin{array}{l} x = \sin\theta\cos\varphi \\
y = \sin\theta\sin\varphi \\ z = \cos\theta \end{array} \right.\longrightarrow
AdS_2 \left\vert\begin{array}{l} x = \cosh\alpha\cos\beta \\
y = \cosh\alpha\sin\beta \\ z = \sinh\alpha \end{array} \right.\,,
\ee et \`a la m\'etrique \be\lb{ADS} d\sigma^2 =  - \cosh^2\alpha\,
d\beta^2 + d\alpha^2 =  - (z^2+1)d\beta^2 + \frac{dz^2}{z^2+1} \,,
\ee qui correspond aux deux premiers termes de (\ref{BR}) avec $b^2
= -1$ (ou avec $b^2 < 0$ quelconque moyennant un changement
d'\'echelle de $z$ et de $\beta$).

Une autre m\'etrique pour $AdS_2$, correspondant \`a (\ref{BR0}),
\be
d\sigma^2 = -r^2 dt^2 + \frac{dr^2}{r^2}\,,
\ee
s'obtient par la param\'etrisation
\be
x+z = r\,, \quad y = tr\,, \quad x-z = \frac1r - t^2r\,.
\ee

Une troisi\`eme m\'etrique, qui ne couvre qu'une partie de $AdS_2$,
r\'esulte de la param\'etrisation \be x = \cosh\alpha\,, \quad y =
\sinh\alpha\sinh\beta\,, \quad z = \sinh\alpha\cosh\beta \ee ($x \ge
1$). Cette m\'etrique, \be d\sigma^2 = - \sinh^2\alpha\, d\beta^2 +
d\alpha^2 = -(x^2-1)d\beta^2 + \frac{dx^2}{x^2-1}\,, \ee correspond
\`a (\ref{BR}) avec $b^2 > 0$. Ceci montre que (malgr\'e l'apparence
de deux horizons en $r = \pm b$), la m\'etrique de Bertotti-Robinson
(\ref{BR}) avec $b^2 > 0$ ne d\'ecrit pas un trou noir, puisqu'une
simple transformation de coordonn\'ees permet d'obtenir son
extension analytique maximale, qui est le produit de la m\'etrique
canonique (\ref{ADS}) sur $AdS_2$ par $S^2$, et est d\'epourvue
d'horizons.

Signalons enfin que l'espace-temps de Bertoti-Robinson peut \^etre
consid\'er\'e comme d\'ecrivant la g\'eom\'etrie pr\`es de l'horizon
\'ev\`enement $r=m$ de la m\'etrique de Reissner-Nordstr\"{o}m
extr\^eme (\ref{RNE}). En posant $r = m+x$ celle-ci s'\'ecrit \be
ds^2 = - \frac{x^2}{(m+x)^2}\,dt^2 + (m+x)^2\left[\frac{dx^2}{x^2} +
d\Omega^2\right]\,, \ee et, dans la limite $x \ll m$, on retrouve
(\ref{BR0}) avec $e = m^{-1}$.

\subsection{Solution de Kerr}
Jusqu'ici, nous avons discut\'e des solutions statiques et \`a sym\'etrie
sph\'erique des \'equations d'Einstein sans source ou avec comme source
un champ \'electro-magn\'etique. La solution de Kerr est une solution
stationnaire, non
plus \`a sym\'etrie sph\'erique, mais \`a sym\'etrie axiale, des \'equations
d'Einstein dans le vide, d\'ependant de deux param\`etres $m$ et $a$. Elle
repr\'esente le champ gravitationnel \`a l'ext\'erieur d'un objet massif en
rotation autour de son axe de sym\'etrie.

On dit qu'une m\'etrique est {\em stationnaire} si, dans un syst\`eme de
coordonn\'ees dit adapt\'e, les composantes du tenseur m\'etrique sont
ind\'ependantes de la coordonn\'ee temps,
\be
ds^2 = g_{\mu\nu}(\vec{x})dx^{\mu}dx^{\nu}\,.
\ee
Dans le cas particulier des m\'etriques statiques (\ref{stat}), les composantes
mixtes $g_{0i}$ sont nulles. Celles-ci sont non nulles dans le cas g\'en\'eral.
L'analyse des propri\'et\'es des trous noirs stationnaires est souvent
facilit\'ee par l'emploi de la param\'etrisation d'Arnowitt-Deser-Misner
(ADM), mise au point initialement pour la formulation canonique et la
quantification de la relativit\'e g\'en\'erale \cite{adm},
\be
ds^2 = -N^2\,dt^2 + h_{ij}(dx^i + N^idt)(dx^j + N^jdt)\,,
\ee
qui d\'ecompose le tenseur m\'etrique d'espace-temps en un scalaire d'espace
$N$ (le ``lapse''), un vecteur d'espace $N^i$ (le ``shift''), et un tenseur
d'espace $h_{ij}$.

La solution de Kerr (1963) s'\'ecrit, en coordonn\'ees dites de
Boyer-Lindquist,
\be
ds^2 - -\frac{\Delta}{h^2}dt^2 + \frac{\rho^2}{\Delta}dr^2 + \rho^2d\theta^2
+ h^2\sin^2\theta\bigg(d\varphi-\frac{2amr}{\rho^2h^2}dt\bigg)^2\,,
\ee
o\`u
\be
\Delta = r^2 - 2mr + a^2\,, \quad \rho^2 = r^2 + a^2 \cos^2\theta\,, \quad
h^2 = r^2 + a^2 + \frac{2mr}{\rho^2}a^2\sin^2\theta\,.
\ee

Pour interpr\'eter les param\`etres $m$ et $a$, consid\'erons la m\'etrique
de Kerr lin\'earis\'ee (\`a grande distance $r \gg m,a$)
\be
ds^2 \simeq - \bigg(1-\frac{2m}{r}\bigg)dt^2 + \bigg(1+\frac{2m}{r}\bigg)dr^2
+ r^2d\theta^2 + r^2\sin^2\theta d\varphi^2 - \frac{4am}{r}\sin^2\theta
d\varphi dt\,.
\ee
En posant
$$ dr = \frac{x^idx^i}{r}\,, \quad r^2\sin^2\theta d\varphi =
(\vec{k}\wedge\vec{x})\cdot d\vec{x}$$
($\vec{k}$ vecteur unitaire de l'axe des $z$), on montre que la d\'eviation
$\gamma_{\mu\nu} = g_{\mu\nu} - \eta_{\mu\nu}$ de la m\'etrique lin\'earis\'ee
par rapport \`a la m\'etrique de Minkowski est donn\'ee par
\be
\gamma_{00} = \frac{2m}r\,, \quad \gamma_{0i} = -\frac{2m}{r^3}
(\vec{a}\wedge\vec{x})^i\,, \quad \gamma_{ij} = \frac{2m}{r^3}x^ix^j
\ee
($\vec{a} = a\vec{k}$),
d'o\`u, en appliquant les formules du 1.2.3, on obtient la masse $M$ et le
moment angulaire $J$ du corps (par exemple une \'etoile) \`a la source de
ce champ de gravitation,
\be
M = m\,, \qquad J = am\,.
\ee

La composante $g_{tt} = -(1-2mr/\rho^2)$ du tenseur m\'etrique
s'annulle sur les {\em surfaces limites} \be\lb{surlim} r = r_{e\pm}
\equiv m \pm \sqrt{m^2-a^2\cos^2\theta}\,, \ee qui sont ferm\'ees si
$a^2 < m^2$. Mais la m\'etrique reste r\'eguli\`ere sur ces
surfaces, o\`u $g_{t\varphi} = -a\sin^2\theta \neq 0$. Par contre,
la m\'etrique de Kerr est singuli\`ere sur les {\em horizons}
$\Delta = 0$, o\`u le carr\'e du lapse $N^2$ s'annulle tandis que la
composante $g_{rr}$ du tenseur m\'etrique diverge, \be r = r_{\pm}
\equiv m \pm \sqrt{m^2-a^2}\,. \ee Le scalaire de Kretschmann reste
r\'egulier sur ces horizons. Enfin, la m\'etrique de Kerr pr\'esente
une vraie {\em singularit\'e} (m\'etrique et de courbure) pour
$\rho^2 = 0$, correspondant \`a \be r=0\,, \quad \theta = \pi/2\,.
\ee On peut montrer par une transformation de coordonn\'ees que
cette singularit\'e n'est pas un point, mais plut\^ot un anneau
(cercle) entourant le disque $r=0$, au travers duquel la m\'etrique
de Kerr peut \^etre prolong\'ee vers les $r$ n\'egatifs. Les
positions relatives des surfaces limites, des horizons et de la
singularit\'e annulaire de la m\'etrique de Kerr sont
sch\'ematis\'ees sur la Figure 7.
\begin{figure}
\centering
\includegraphics[width=8cm,angle=0]{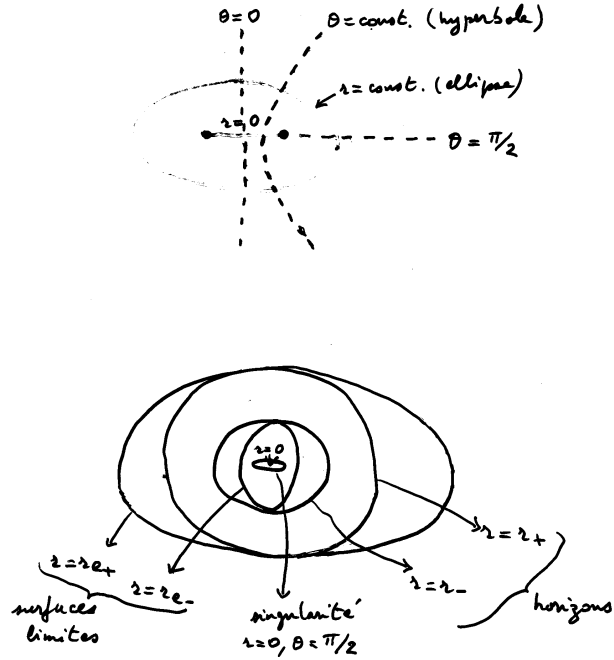}
\caption{Surfaces limites, horizons et singularit\'e de la
m\'etrique de Kerr dans le cas $a^2<m^2$} \label{fig7}
\end{figure}

La m\'etrique de Kerr peut \^etre prolong\'ee \`a la Kruskal \`a
travers ses deux horizons (dans le cas $a^2<m^2$), ou son horizon
double (dans le cas extr\^eme $a^2=m^2$). Le diagramme de Penrose de
son extension analytique maximales est donn\'e dans la Figure 8 pour
le cas $a^2<m^2$. Ce diagramme comporte trois types de r\'egions,
empil\'ees sous la forme d'une tour infinie:
\begin{itemize}
\item r\'egions $I$: $r \in [r_+, \infty[$, o\`u la signature est $(-+++)$
(une de ces r\'egions est la r\'egion de l'observateur en Scri$_+$);
\item r\'egions $II$: $r \in [r_-, r_+]$, o\`u la signature est $(+-++)$;
\item r\'egions $III $: $r \in ]-\infty,r_-]$, o\`u la signature est $(-+++)$
(ces r\'egions contiennent le disque $r=0$ et la singularit\'e annulaire
$r = 0$, $\theta = \pi/2$).
\end{itemize}
Carter a montr\'e en 1968 \cite{carter} que le mouvement g\'eod\'esique dans la
m\'etrique de Kerr est compl\`etement int\'egrable. Les g\'eod\'esiques non
radiales de genre temps peuvent aller de la r\'egion $I$ \`a la r\'egion
$III'$ en traversant successivement les deux horizons $r = r_+$, $r = r_-$,
puis le disque $r = 0$, seules les g\'eod\'esiques radiales de genre temps se
terminant sur la singularit\'e annulaire. Le diagramme de Penrose pour le cas
extr\^eme $a^2 = m^2$  s'obtient \`a partir du cas $a^2<m^2$ en supprimant
les r\'egions $II$ et $II'$. Enfin, dans le cas $a^2>m^2$, la singularit\'e
annulaire est nue.
\begin{figure}
\centering
\includegraphics[width=8cm,angle=0]{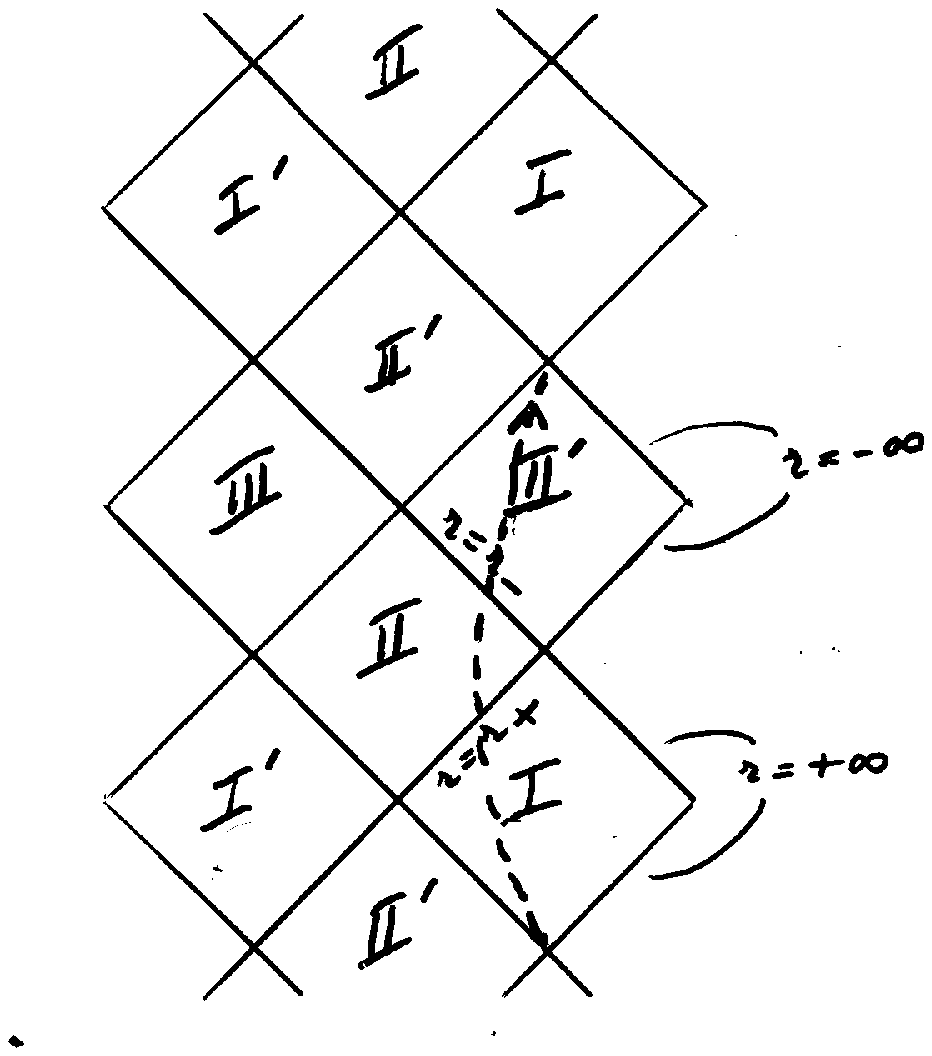}
\caption{Diagramme de Penrose de Kerr dans le cas $a^2 < m^2$}
\label{fig8}
\end{figure}

Une propri\'et\'e nouvelle de l'espace-temps de Kerr (par rapport
\`a ceux de Schwarzschild ou de Reissner-Nordstr\"{o}m) est
l'existence d'une {\em ergosph\`ere}, qui est la r\'egion $r_{e-} <
r < r_{e+}$, comprise entre les surfaces limites (\ref{surlim}),
o\`u $g_{tt} = - N^2 + g_{\varphi\varphi}(N^{\varphi})^2 > 0$. Le
temps propre n'\'etant pas d\'efini dans cette r\'egion, puisque
$d\tau^2 = -g_{tt}dt^2 < 0$, un observateur ne peut pas rester
statique (au repos) dans l'ergosph\`ere. Mais l'observateur a la
possibilit\'e d'\^etre stationnaire, c'est-\`a dire de tourner dans
l'ergosph\`ere \`a distance constante de l'horizon avec une vitesse
angulaire comprise entre les limites \be 0 < \frac{2amr}{\rho^2h^2}
- \frac{\sqrt{\Delta}}{h^2|\sin\theta|} < \Omega <
\frac{2amr}{\rho^2h^2} + \frac{\sqrt{\Delta}}{h^2|\sin\theta|}\,.
\ee Ces limites s'obtiennent en \'ecrivant \be d\tau^2 = N^2dt^2 -
g_{\varphi\varphi}(d\varphi + N^{\varphi}dt)^2 > 0\,, \ee et en
rempla\c{c}ant $d\varphi = \Omega dt$, d'o\`u \be |\Omega +
N^{\varphi}| < \sqrt{\frac{N^2}{g_{\varphi\varphi}}} < |N^{\varphi}|
\,. \ee

Dans l'ergosph\`ere, l'\'energie est de genre espace, donc peut \^etre
n\'egative. Il en r\'esulte la possibilit\'e d'extraire de l'\'energie du
trou noir de Kerr par le {\em processus de Penrose}. Ce processus classique
peut \^etre sch\'ematis\'e en trois \'etapes:
\begin{enumerate}
\item Une particule d'\'epreuve d'\'energie $E_0$ tombe \`a l'int\'erieur de
l'ergosph\`ere.
\item
Avant d'atteindre l'horizon, elle se d\'esint\`egre en deux fragments,
$E_0 = E_1 + E_2$, avec $E_1 < 0$ et $E_2 > 0$.
\item Le fragment d'\'energie n\'egative tombe dans le trou noir (avec un
moment angulaire oppos\'e \`a $J = Ma$), tandis que le fragment d'\'energie
positive repart \`a l'infini avec une \'energie $E_2 = E_0 - E_1 > E_0$.
\end{enumerate}
L'\'energie maximum qu'on peut ainsi extraire du trou noir est \linebreak
$m[1-\sqrt{(1+\sqrt{1-a^2/m^2})/2}]$. Le trou noir r\'esiduel est statique.

Pour terminer, signalons que la th\'eorie d'Einstein-Maxwell admet \'egalement
une solution stationnaire, la solution de Kerr-Newman, correspondant \`a des
trous noirs charg\'es en rotation. Ces trous noirs ont une masse $M$, un
moment angulaire $J$ et une charge \'electrique $Q$. On montre par ailleurs
que ce sont les seules observables d'un trou noir pouvant \^etre mesur\'ees
par un observateur ext\'erieur.

\subsection{Thermodynamique des trous noirs}
Bien qu'il s'agisse d'un processus quantique, nous mentionnerons
rapidement le {\em rayonnement de Hawking}, qui joue un r\^ole
important dans la physique des trous noirs \`a l'\'echelle
cosmologique. Ce processus consiste en la cr\'eation de paires
particule-antiparticule dans le champ graviationnel ext\'erieur du
trou noir, au niveau de l'horizon. L'antiparticule tombe dans le
trou noir sous la forme d'une particule d'\'energie n\'egative, qui
diminue donc la masse du trou noir, tandis que la particule (par
exemple un photon) d'\'energie positive part \`a l'infini, sous la
forme de rayonnement.

Le calcul semi-classique effectu\'e par Hawking \cite{haw} dans le
cas du trou noir de Schwarzschild montre que le spectre de
l'\'energie ainsi rayonn\'ee est celui du rayonnement d'un corps
noir, \`a la temp\'erature de Hawking \be T_H = \frac{\hbar
c^3}k\frac1{8\pi GM} = 6\times10^{-8} \frac{M_{\odot}}M\, ^oK\,. \ee
Les trous noirs de temp\'erature sup\'erieure \`a la temp\'erature
du rayonnement de fond cosmologique, $T_H > 2.7 ^oK$, soit $M <
2\times10^{-8} M_{\odot} \simeq M_{lune}$ rayonnent et perdent
graduellement de leur masse. Les trous noirs plus lourds sont
stables vis-\`a-vis du processus de Hawking.

L'\'energie rayonn\'ee par l'horizon du trou noir peut \^etre calcul\'ee \`a
l'aide de la loi de Stefan:
\be
\frac{d(Mc^2)}{dt} = \sigma T_H^4 \times 4\pi r_h^2 \propto M^{-2}\,.
\ee
Il en r\'esulte la dur\'ee de vie d'un trou noir,
\be
\tau \approx 10^{58} \bigg(\frac{M}{M_{\odot}}\bigg)^3 \mbox{\rm ans}\,.
\ee
Un trou noir primordial (cr\'e\'e aux premiers instants de l'Univers) doit,
pour \^etre observable aujourd'hui, avoir eu une dur\'ee de vie sup\'erieure
\`a, ou de l'ordre de, l'\^age de l'Univers $10^{10}$ ans, ce qui donne pour
les mini-trous noirs primordiaux la limite inf\'erieure de masse initiale
$M > 10^{-16} M_{\odot}$ ($10^{14}$ kg).

Les trous noirs charg\'es en rotation sont donc des corps ob\'eissant aux {\em
lois de la thermodynamique}. La premi\`ere loi, qui \'etablit une relation
entre les variations des observables du trou noir, est particuli\`erement
importante:

\bigskip
\noindent Premi\`ere loi: {\em Les variations des param\`etres du trou noir
sont reli\'ees par}
\be
\delta M = T_H\delta S + \Omega_h
\delta J + V_h \delta Q\,,
\ee
o\`u
\be
S = \frac{kc^3}{\hbar} \frac{4\pi r_h^2}{4G}
\ee
est l'entropie du trou noir \cite{bek}, $\Omega_h = - N^{\varphi}(r_h)$ est la
vitesse angulaire de l'horizon, et $V_h = A_0(r_h)$ est le potentiel
\'electrique de l'horizon.

La deuxi\`eme loi, $\delta S \ge 0$, entra\^ine qu'au niveau classique (donc
abstraction faite du rayonnement de Hawking) la somme des aires des horizons
des trous noirs, donc l'\'energie totale contenue dans les trous noirs de
l'Univers, ne peut que cro\^{\i}tre, par accr\'etion de mati\`ere par les
trous noirs. Cette croissance est \'evidemment contrebalanc\'ee par
l'\'evaporation de Hawking.

\section{Mod\`eles sigma}
\setcounter{equation}{0} Apr\`es cette revue de diff\'erentes
solutions stationnaires des \'equations d'Einstein et
d'Einstein-Maxwell, nous allons montrer comment ces solutions
peuvent \^etre reli\'ees entre elles par des transformations
appartenant \`a un groupe de sym\'etrie cach\'e du syst\`eme
d'\'equations des champs.

\subsection{Vecteurs de Killing}

On dit qu'une transformation de coordonn\'ees $ x \to x' $ est une
{\em isom\'etrie} de la m\'etrique $g_{\mu\nu}(x)$ si \be\lb{iso}
g'_{\mu\nu}(x') \equiv \frac{\partial x^{\alpha}}{\partial
x'_{\mu}}\frac{\partial x^{\beta}}{\partial
x'_{\nu}}g_{\alpha\beta}(x) = g_{\mu\nu}(x')\,. \ee Dans le cas
d'une transformation de coordonn\'ees infinit\'esimale,
\be x^{'\mu}
= x^{\mu} + \varepsilon\xi^{\mu}(x) \qquad (|\varepsilon| \ll
1)\,,\ee
les \'el\'ements de la matrice de transformation de
coordonn\'ees se r\'eduisent \`a
\be \frac{\partial
x^{\alpha}}{\partial x'_{\mu}} = \delta^{\alpha}_{\mu} -
\varepsilon\partial_{\mu}\xi^{\alpha}\,, \ee et la d\'efinition  de
l'isom\'etrie (\ref{iso}) se r\'eduit \`a
\be
\partial_{\mu}\xi^{\lambda}g_{\lambda\nu} +
\partial_{\nu}\xi^{\lambda}g_{\mu\lambda} +
\partial_{\lambda}g_{\mu\nu}\xi^{\lambda}
= 0\,,\ee
qui se traduit en termes de d\' eriv\' ees covariantes par
\be\lb{kil} D_{\mu}\xi_{\nu} + D_{\nu}\xi_{\mu} = 0\,. \ee

On dit que $\xi_{\mu}(x)$ est un champ de {\em vecteurs de Killing}
de la m\'etrique $g_{\mu\nu}(x)$, et on associe \`a ce champ de
vecteurs l'op\'erateur d\'erivation $$J[\xi](x) \equiv
\xi^{\mu}(x)\partial_{\mu},$$ qui est proportionnel au
g\'en\'erateur de l'isom\'etrie infinit\'esimale
\be
\left[\varepsilon J[\xi](x), x^{\mu}\right] =
\varepsilon\xi^{\mu}(x)\,,\ee
que, par abus de langage, nous appellerons aussi vecteur de Killing.

\subsection{R\'eduction des \'equations d'Einstein stationnaires}
La notion de vecteur de Killing nous permet de d\'efinir de
fa\c{c}on covariante une m\'etrique stationnaire comme une
m\'etrique poss\'edant un champ de vecteurs de Killing
$\xi_{\mu}(x)$ de genre temps, \be |\xi|^2 \equiv
g_{\mu\nu}\xi^{\mu}\xi^{\nu} < 0\,. \ee On retrouve la d\'efinition
des m\'etriques stationnaires donn\'ee pr\'ec\'edemment dans le cas
o\`u $\xi = \partial_t$. En effet, dans ce cas les transformations
infinit\'esimales engendr\'ees par $\xi$ sont les translations dans
la variable $t$, qui est bien le temps car $g_{tt} < 0$.

Consid\'erons la gravitation d'Einstein sans source avec un vecteur
de Killing  de genre temps. De m\^eme qu'on peut d\'ecomposer le champ
\'electro-magn\'etique $F_{\mu\nu}$ en un vecteur d'espace
$F_{0i} = E_i$ (champ \'electrique) et un pseudo-vecteur d'espace
$F_{ij} = \epsilon_{ijk}B^k$ (champ magn\'etique), on peut toujours
d\'ecomposer
le tenseur m\'etrique $g_{\mu\nu}$ (tenseur sym\'etrique $4\times4$, donc
10 composantes) en un scalaire d'espace $g_{00} = -f$ (potentiel
gravi\'electrique), un vecteur d'espace $g_{0i} = -f\A_i$ ($\A_i$ potentiel
vecteur gravimagn\'etique) et un tenseur d'espace $g_{ij} =
(f^{-1}h_{ij}-f\A_i\A_j)dx^idx^j$ (10 = 1 + 3 + 6). Sous forme compacte,
\begin{equation}\label{st4}
ds^2 =  -f(dt + \A_idx^i)^2 + f^{-1}h_{ij}dx^idx^j\,,
\end{equation}
o\`u $i,j = 1,2,3$, et les champs $f$, $\A_i$ et $h_{ij}$
d\'ependent seulement des coordonn\'ees d'espace $x^k$. Cette
d\'ecomposition (\`a ne pas confondre avec la d\'ecomposition ADM)
est souvent appel\'ee ``r\'eduction de Kaluza-Klein'' de quatre \`a
trois dimensions (par analogie avec la r\'eduction de cinq \`a
quatre dimensions).

Les dix \'equations d'Einstein $R_{\mu}^{\nu} = 0$ se d\'ecomposent de m\^eme
en \'equations scalaire d'espace, vectorielle d'espace et tensorielle d'espace.
Commen\c{c}ons par l'\'equation vectorielle $R_0^i = 0$. Elle s'\'ecrit
\be\lb{R0i}
D_j(f^2\F^{ij}) = \frac1{\sqrt{|h|}}\partial_j(\sqrt{|h|}f^2\F^{ij}) = 0\,,
\ee
o\`u
\be
\F_{ij} \equiv D_i\A_j - D_j\A_i = \partial_i\A_j - \partial_j\A_i\,,
\ee
$D_i$ est la d\'eriv\'ee covariante \`a trois dimensions d\'efinie par rapport
\`a la m\'etrique r\'eduite $h_{ij}$, les indices sont \'elev\'es par la
m\'etrique r\'eduite inverse $h^{ij}$, et $h =$ det$(h_{ij})$. L'\'equation
(\ref{R0i}), analogue (au $\sqrt{|h|}f^2$ pr\`es) \`a l'\'equation de
Maxwell sans source $\partial_jF^{ij}=0$ ($\nabla\wedge\vec{B} = 0$),
peut \^etre \'ecrite symboliquement
\be
\nabla\wedge(f^2\nabla\wedge\vec\A) = 0\,,
\ee
o\`u $\nabla\wedge$ est l'op\'erateur rotationnel covariant (voir \'equation
(\ref{om2})). Elle est r\'esolue par
\be\lb{om1}
\nabla\wedge\vec{\A} = f^{-2}\nabla\omega\,,
\end{equation}
ou explicitement
\be\lb{om2}
\frac1{\sqrt{|h|}}\epsilon^{ijk}\partial_j\A_k = f^{-2}h^{il}
\partial_l\omega\,.
\ee
Cette {\em \'equation de dualit\'e} nous permet de remplacer le potentiel
vecteur gravimagn\'etique $\A$ par le potentiel scalaire de {\em twist}
$\omega$ qui, par sa d\'efinition (\ref{om1}), v\'erifie l'\'equation
\begin{equation}\label{om}
\nabla(f^{-2}\nabla\omega)=0\,.
\end{equation}
L'\'equation scalaire $R_{00} = 0$ s'\'ecrit
\be\lb{f}
\nabla(f^{-2}\nabla f) = -f^{-3}[(\nabla f)^2 + (\nabla\omega)^2]\,.
\end{equation}
Enfin, l'\'equation tensorielle $R^{ij} = 0$ s'\'ecrit
\begin{equation}\label{eine4}
R_{(3)ij} = \frac1{2f^2} \bigg(\partial_i f\partial_j f +
\partial_i\omega
\partial_j\omega\bigg) \,,
\end{equation}
o\`u $R_{(3)ij}$ est le tenseur de Ricci pour la m\'etrique r\'eduite
$h_{ij}$.

Les trois \'equations (\ref{om}), (\ref{f}) et (\ref{eine4}) d\'erivent de
l'action r\'eduite
\begin{equation}\lb{redac}
S_{(3)} = \int d^3x\sqrt{|h|}\bigg[-R_{(3)} +
G_{AB}(X)\partial_iX^A
\partial_jX^Bh^{ij}\bigg]\,,
\end{equation}
o\`u $G_{AB}$ est la m\'etrique de l'espace cible $\cal T$ (espace des
potentiels $X^A$)
\begin{equation}\label{tare4}
dS^2 \equiv G_{AB}dX^AdX^B = \frac1{2f^2}(df^2+d\omega^2)\,.
\end{equation}
Le deuxi\`eme terme de l'action effective (\ref{redac})
caract\'erise un mod\`ele sigma (non-lin\'eaire). La th\'eorie
d'Einstein sans source \`a quatre dimensions (E4) avec un vecteur de
Killing se r\'eduit donc \`a un mod\`ele sigma coupl\'e \`a la
gravitation \`a trois dimensions\footnote{Une bonne introduction aux
mod\`eles sigma autogravitants peut \^etre trouv\'ee dans
\cite{BGM}.}.

Quelles sont les sym\'etries manifestes de l'espace cible? L'invariance de
E4 par les diff\'eomorphismes \`a quatre dimensions a \'et\'e bris\'ee par la
r\'eduction \`a trois dimensions. Il subsiste, outre les diff\'eomorphismes
\`a trois dimensions et les translations dans le temps, les dilatations
combin\'ees
$$t \to \alpha^{-1} t\,, \quad f \to \alpha^2 f\,, \quad
\omega \to \alpha^{2}\omega\,,$$ qui laissent invariant l'ansatz
m\'etrique (\ref{st4}). Ces dilatations (qu'on peut, par analogie
avec l'\'electromagn\'etisme, qualifier de ``transformations de
jauge'') sont engendr\'ees par le vecteur de Killing de {\cal T}
\begin{equation}
M = 2(f\partial_f + \omega\partial_{\omega})\,.
\end{equation}
L'ansatz (\ref{st4}) est \'egalement invariant sous les transformations de
jauge $t \to t + \phi(\vec{x})$, $\A_i(\vec{x}) \to \A_i(\vec{x}) - \partial_i
\phi(\vec{x})$, qui ne modifient pas (voir (\ref{om1})) les twists
$\omega(\vec{x})$. Mais il est aussi \'evidemment invariant sous les
translations de $\omega$, engendr\'ees par le vecteur de Killing
\begin{equation}
N = \partial_{\omega}\,.
\end{equation}

En plus de ces sym\'etries manifestes, la r\'esolution des \'equations de
Killing pour la m\'etrique (\ref{tare4}) montre qu'elle admet \'egalement
une {\em sym\'etrie cach\'ee}, engendr\'ee par le vecteur de Killing
\begin{equation}
L = (\omega^2-f^2)\partial_{\omega} + 2\omega f\partial_f\,.
\end{equation}
Il est facile de v\'erifier que les trois vecteurs de Killing $L$, $M$
et $N$ engendrent l'alg\`ebre de Lie associ\'ee au groupe $SL(2,R)$:
\begin{eqnarray}\lb{sl2r}
\left[M,N\right] &=& -2N \,,\nonumber\\
\left[M,L\right] &=& 2L \,,\\
\left[N,L\right] &=& M \,.\nonumber
\end{eqnarray}
On montre que  $\cal T$ est l'{\em espace sym\'etrique} (ou {\em coset})
$SL(2,R)/SO(2)$\footnote{\lb{EinEM}Le groupe $SL(2,R)$ est localement isomorphe
au groupe $SU(1,1)$. Il en r\'esulte que $\cal T$ peut aussi \^etre
identifi\'e au coset $SU(1,1)/S[U(1)\times U(1)]$.}, o\`u $H = SO(2)$ est le
sous-groupe d'isotropie (groupe des transformations pr\'eservant un point
donn\'e $X_0^A$ de \T) du groupe $G = SL(2,R)$. La dimension de cet espace
est 3 (dim[$SL(2,R)$]) - 1 (dim[$SO(2)$]) = 2.

Un exemple g\'eom\'etrique simple d'espace sym\'etrique $G/H$ est celui de la
sph\`ere $S^2$. La sph\`ere (qu'on peut se repr\'esenter comme plong\'ee
dans l'espace eucliden \`a trois dimensions) est globalement invariante par
les rotations qui forment le groupe $SO(3)$. Parmi ces transformations,
celles qui pr\'eservent un point donn\'e sont les rotations autour de l'axe
passant par ce point (et le point antipode), qui forment le sous-groupe
d'isotropie $SO(2)$.  Et la sph\`ere $S^2$ est l'espace
sym\'etrique $SO(3)/SO(2)$. Un autre exemple d'espace sym\'etrique est fourni
par $AdS_2 = SO(2,1)/SO(1,1)$.

\subsection{Repr\'esentant matriciel}

Au point de coordonn\'ees $(f,\omega)$ du coset $SL(2,R)/SO(2)$, on peut
associer la matrice $2\times2$ r\'eelle, sym\'etrique, unimodulaire (de
d\'eterminant unit\'e), dite {\em repr\'esentant matriciel}
\begin{equation}
{\cal M}[f,\omega] = \left(\begin{array}{cc}
f + f^{-1}\omega^2 & -f^{-1}\omega \\
-f^{-1}\omega &  f^{-1}
\end{array}\right)\,,
\end{equation}
qui permet d'\'ecrire la m\'etrique (\ref{tare4}) sous la forme
\begin{equation}\label{tarmet}
dS^2 = \frac14{\rm Tr}({\cal M}^{-1}d{\cal M} {\cal M}^{-1}d{\cal
M})\,,
\end{equation}
tandis que les \'equations (\ref{om}), (\ref{f}) et (\ref{eine4})
s'\'ecrivent
\begin{eqnarray}
\nabla({\cal M}^{-1}\nabla{\cal M}) &=& 0\,, \label{divJ}\\
R_{(3)ij} &=& \frac14{\rm Tr}({\cal M}^{-1}\partial_i{\cal M}
{\cal M}^{-1}\partial_j{\cal M})\,. \label{Rij}
\end{eqnarray}

Il est clair que ces \'equations sont invariantes ($h^{'}_{ij}= h_{ij}$) sous
les transformations globales du groupe $G = SL(2,R)$ agissant
bilin\'eairement sur $\M$
\begin{equation}\label{transf}
{\cal M}(\x) \to {\cal M}'(\x) = P{\cal M}(\x)P^T \,,
\end{equation}
o\`u  $P$ est une matrice r\'eelle unimodulaire, et $P^T$ la matrice
transpos\'ee. Donc, en faisant agir sur une solution des \'equations
d'Einstein, \'ecrite
sous la forme $(h_{ij}(\x),\M(\x))$, une transformation $P \in G$ on obtient
une nouvelle solution des \'equations d'Einstein.

Les transformations infinit\'esimales sont de la forme $P = 1 +
\varepsilon j$, o\`u $j$ appartient \`a la repr\'esentation
matricielle de l'alg\`ebre de Lie $sl(2,R)]$. La condition
det$\,P=1$ implique que la matrice $j$ (matrice r\'eelle $2\times2$)
est de trace nulle. On peut v\'erifier que ceci entra\^ine \be j^T =
KjK\,, \qquad K = \left(\begin{array}{cc} 0 & 1 \\ -1 & 0
\end{array}\right)\,,
\ee avec $K^2 = -1$. L'action d'une telle transformation
infinit\'esimale sur $\M$ conduit \`a $\M' = \M + \delta\M$ avec \be
\delta(\M K) = \left[\varepsilon j,\M K\right]\,. \ee Par
construction, cette transformation est une isom\'etrie
infinit\'esimale de la m\'etrique (\ref{tare4}), donc a aussi \be
\delta(\M K) = \left[\varepsilon J,\M K\right]\,, \ee o\`u $J$ est
le correspondant de $j$ dans l'alg\`ebre des op\'erateurs
d\'erivations (vecteurs de Killing de $\T$). On peut v\'erifier que
les repr\'esentants matriciels des op\'erateurs $L$, $M$ et $N$, \be
l = \left(\begin{array}{cc} 0 & 0
\\ 1 & 0 \end{array}\right)\,, \quad m = \left(\begin{array}{cc} 1 &
0 \\ 0 & -1 \end{array}\right)\,, \quad n = \left(\begin{array}{cc}
0 & -1 \\ 0 & 0 \end{array}\right) \ee ob\'eissent aux relations de
commutation de l'alg\`ebre (\ref{sl2r}).

Quelles sont les transformations qui pr\'eservent le comportement
asymptotique? L'espace-temps \`a quatre dimensions de m\'etrique
(\ref{st4}) est asymptotiquement plat si l'espace r\'eduit \`a trois
dimensions est asymptotiquement plat
($R_{(3)ij}(\infty)=0$)\footnote{\lb{Rij3} A trois dimensions, le
tenseur de Riemann et le tenseur de Ricci ont le m\^eme nombre (6)
de composantes ind\'ependantes, donc $R_{ij}=0 \Rightarrow
{R^i}_{jkl} = 0$.\lb{r3}} et si $f(\infty) = f_0$ (constante) et
$\A_i(\vec{x}) \to 0$ suffisamment vite quand $|\vec{x}| \to
\infty$, conduisant \`a $\omega(\infty) = \omega_0$ (constante). Les
transformations qui pr\'eservent le comportement asymptotique sont
donc celles qui laissent invariant un point donn\'e
$(f_0,\omega_0)$. En utilisant les transformations de jauge $M$ et
$N$ on peut toujours se ramener (``fixer la jauge'')  aux conditions
\be\lb{as} f(\infty)=1\,, \quad \omega(\infty) = 0 \quad
\longleftrightarrow \quad \M(\infty)=\left(\begin{array}{cc} 1 & 0
\\ 0 & 1 \end{array}\right)\,. \ee Il est facile de voir que ces
conditions sont pr\'eserv\'ees par la transformation
infinit\'esimale \be L+N = (1 - f^2 + \omega^2)\partial_{\omega} +
2\omega f\partial_f \quad \longleftrightarrow \quad l+n =
\left(\begin{array}{cc} 0 & -1 \\ 1 & 0 \end{array}\right)\,, \ee
qui engendre le groupe d'isotropie $H=SO(2)$.

L'action (\ref{transf}) d'une transformation finie de $SO(2)$,
\begin{equation}\lb{NUT}
P = \e^{\ds\alpha(l+n)} = \left(\begin{array}{cc}
\cos\alpha & -\sin\alpha \\ \sin\alpha & \cos\alpha
\end{array}\right)\,.
\end{equation}
sur la solution de Schwarzschild $f_0 = 1 - 2m_0/r_0$, $\omega_0 =
0$, conduit \`a une solution $(f,\omega)$ avec un potentiel de twist
monopolaire (\`a sym\'etrie sph\'erique) non nul, qui donne apr\`es
dualisation inverse (r\'esolution de l'\'equation (\ref{om1})) un
potentiel gravimagn\'etique monopolaire $\A_{\varphi} =
2\ell\cos\theta$. La solution obtenue est la solution de Taub-NUT
\ba\lb{nut} ds^2 &=& -f(r)(dt+2\ell\cos\theta\,d\varphi)^2 +
f^{-1}(r)\,dr^2 + (r^2 + \ell^2)\,
d\Omega^2\,, \\
&& \qquad f(r) = \frac{r^2-2mr-\ell^2}{(r^2+\ell^2)^2}\,, \nn \ea
o\`u $r = r_0 - 2m_0\sin^2\alpha$, et la masse et la charge NUT sont
reli\'ees \`a la masse de Schwarzschild par \be m =
m_0\cos2\alpha\,, \quad \ell = m_0\sin2\alpha \qquad (m^2 + \ell^2 =
m_0^2)\,. \ee Cette solution est une curiosit\'e math\'ematique sans
signification physique car, comme le monop\^ole magn\'etique de
Dirac, elle est singuli\`ere le long d'une ``corde''\footnote{En
fait, la m\'etrique (\ref{nut}) est singuli\`ere sur les deux
demi-droites $\theta = 0$ et $\theta = \pi$, mais on peut se ramener
\`a une seule demi-droite singuli\`ere par la transformation de
coordonn\'ees $dt \to dt + 2\ell d\varphi$.} allant de l'origine
$r=0$ \`a l'infini, et n'est donc pas vraiment asymptotiquement
plate.

\subsection{Solutions d\'ependant d'un seul potentiel}
Revenons \`a l'action r\'eduite, \'ecrite pour un espace cible $\T$
quelconque,
\be\lb{act3}
S_{(3)} = \int d^3x\sqrt{|h|}\bigg[-R_{(3)} +
G_{AB}(X)\nabla X^A \cdot \nabla X^B\bigg]\,.
\ee
Les \'equations d'Euler-Lagrange pour les champs $X^A(\vec{x})$ s'\'ecrivent
\be\lb{el3}
\nabla_{(h)}^2X^A + \Gamma^A_{BC}(X)\nabla X^B\nabla X^C = 0\,,
\ee
o\`u $\nabla_{(h)}^2$ est le Laplacien covariant pour la m\'etrique r\'eduite
$h_{ij}$. Cherchons \`a r\'esoudre ces \'equations dans le cas o\`u les $X^A$
d\'ependent d'un seul potentiel scalaire $\sigma(x)$,
\be\lb{1pot}
X^A(\x) = X^A[\sigma(\x)]\,.
\ee
 En reportant
(\ref{1pot}) dans (\ref{el3}), on obtient
\be
[\ddot{X}^A + \Gamma^A_{BC}\dot{X}^B\dot{X}^C](\nabla\sigma)^2 + \dot{X}^A
\nabla^2_{(h)}\sigma = 0\,,
\ee
Cette \'equation \'etant du type $\nabla^2_{(h)}\sigma = f(\sigma)
(\nabla\sigma)^2$, on peut
par une reparam\'etrisation $\sigma \to \sigma'[\sigma]$ se ramener au
syst\`eme
\ba
&& \nabla^2_{(h)}\sigma = 0\,, \lb{harmpot}\\
&& \ddot{X}^A + \Gamma^A_{BC}\dot{X}^B\dot{X}^C = 0\, \lb{geo3}
\ea
On reconnait dans cette derni\`ere \'equation l'\'equation des g\'eod\'esiques
de l'espace cible.

En termes du repr\'esentant matriciel $\M[\sigma(\x)]$, l'\'equation
g\'eod\'esique est l'\'equation (\ref{divJ}) qui s'\'ecrit, compte tenu de
(\ref{harmpot}),
\be
\nabla(\M^{-1}\nabla\M) = \frac{d}{d\sigma}\left[\M^{-1}\frac{d\M}{d\sigma}
\right](\nabla\sigma)^2 = 0\,.
\ee
Cette \'equation diff\'erentielle matricielle s'int\`egre par
\begin{equation}\label{etaA}
\M(\x) = \eta{\rm e}^{A\sigma(\x)}\,,
\end{equation}
o\`u $\sigma(\x)$ est une fonction $h$-harmonique, et les matrices constantes
$A \in {\rm Lie}(G)$ et $\eta$ doivent remplir certaines conditions pour que
(\ref{etaA}) soit bien un repr\'esentant matriciel du coset $G/H$. Par
exemple, dans le cas de E4, les conditions d'unimodularit\'e et de sym\'etrie
de $\M$ entra\^{\i}nent que $\eta$ est unimodulaire et sym\'etrique, et
\begin{equation}\label{condA}
{\rm Tr}A = 0\,, \quad A^T\eta-\eta A = K\{A,K\eta\} = 0
\end{equation}
($A$ anticommute avec $K\eta$). Enfin, dans le cas o\`u $\sigma(\infty)=0$,
alors $\eta = \M_{\infty}$.

Il ne nous reste plus qu'\`a r\'esoudre les
\'equations (\ref{harmpot}) et (\ref{Rij}), qui s'\'ecrivent
\ba
&& \partial_i(\sqrt{|h|}h^{ij}\partial_j\sigma) = 0\,, \lb{harmpot2} \\
&& R_{(3)ij} = \frac{1}{4}{\rm Tr}(A^2)\partial_i\sigma\partial_j\sigma\,.
\ea
Nous avons donc ramen\'e le probl\`eme \`a celui d'un champ scalaire de masse
nulle coupl\'e \`a la gravitation \`a trois dimensions avec une constante de
couplage proportionnelle \`a ${\rm Tr}(A^2)$. Cette m\'ethode peut par
exemple \^etre appliqu\'ee \`a la construction de solutions \`a sym\'etrie
sph\'erique.

Une autre application int\'eressante est la construction de {\em solutions
multi-centres}. Le long d'une g\'eod\'esique de $\T$ param\'etris\'ee par
$\sigma(\x)$, l'\'el\'ement de longueur (\ref{tarmet}) s'\'ecrit
\begin{equation}
dl^2=\frac{1}{4}{\rm Tr}(A^2)\,d\sigma^2\,.
\ee
Donc le genre de la g\'eod\'esique d\'epend du signe de ${\rm Tr}(A^2)$.
Dans le cas de E4, la m\'etrique (\ref{tare4}) n'admet que des g\'eod\'esiques
de genre espcae. Plus g\'en\'eralement, si la m\'etrique de l'espace
cible n'est pas d\'efinie positive, elle admet aussi des g\'eod\'esiques de
genre temps, ainsi que des g\'eod\'esiques de genre lumi\`ere
caract\'eris\'ees par
\begin{equation}\label{anti}
{\rm Tr}(A^2) = 0\,.
\end{equation}
Alors, $R_{(3)ij}=0$, donc l'espace r\'eduit \`a trois dimensions est plat
(voir la note \ref{Rij3}). Dans le cas o\`u cet espace est
l'espace euclidien R$^3$, l'\'equation (\ref{harmpot2}) est l'\'equation
de Laplace ordinaire, admettant les solutions multi-centres
\begin{equation}
\sigma(\vec{x}) = \sum_{i}\frac{m_i}{|\vec{x}-\vec{x_i}|}
\,.
\end{equation}
Le probl\`eme de la construction de solutions multi-centres pour le
mod\`ele sigma autogravitant consid\'er\'e se ram\`ene donc \`a la
construction de matrices $A$ remplissant les conditions
mentionn\'ees apr\`es (\ref{etaA}) et la condition suppl\'ementaire
(\ref{anti}). Cette m\'ethode a permis de retrouver les solutions de
Majumdar-Papapetrou (\ref{MP}) de la th\'eorie d'Einstein-Maxwell
\cite{bps}, et de construire les solutions multi-centres de la
th\'eorie d'Einstein \`a cinq dimensions \cite{spat} (voir
sous-section 3.8). Elle a \'egalement \'et\'e g\'en\'eralis\'ee au
cas de solutions multi-centres d\'ependant de plusieurs potentiels
\cite{spat,bps}.

\subsection{Einstein-Maxwell}
Dans le cas de la th\'eorie d'Einstein-Maxwell avec un vecteur de Killing de
genre temps (EM4) , le tenseur m\'etrique peut \`a nouveau \^etre r\'eduit
de quatre \`a trois dimensions par
\begin{equation}\lb{st41}
ds^2 =  -f(dt + \A_idx^i)^2 + f^{-1}h_{ij}dx^idx^j\,.
\end{equation}
Le champ \'electrostatique $E_i = - F_{i0}$ et le champ
magn\'etostatique  sans source \be\lb{magn} B_i = F^*_{i0} =
\frac{\sqrt{|h|}}{2f}\epsilon_{ijk}F^{jk} \ee d\'erivent de
potentiels scalaires $v$ et $u$; \be\lb{vu} \vec{E} = - \nabla v\,, \quad
\vec{B} = \nabla u\,. \ee La composante vectorielle d'espace des
\'equations d'Einstein avec source (\ref{einM})  \be R_0^i =
2F_{0j}F^{ij} = f\left[\nabla\wedge(u\nabla v - v\nabla u)\right]^i
\ee s'int\`egre par \be\lb{chi} f^2\nabla\wedge\vec{\A} + 2(u\nabla
v - v \nabla u) = \nabla\chi\,, \ee o\`u le potentiel de twist
$\chi$ g\'en\'eralise le $\omega$ du cas sans  source (\'equation
(\ref{om1})).

Les quatre potentiels scalaires $f$, $\chi$, $v$ et $u$ peuvent
\^etre regroup\'es en deux potentiels complexes, dits potentiels
d'Ernst, \be\lb{Epsi} \E = f + i\chi - \ol{\psi}\psi \,, \quad \psi = v +
iu\,.  \ee On montre alors que les \'equations stationnaires
d'Einstein-Maxwell restantes (ainsi que  les conditions
d'int\'egrabilit\'e pour les \'equations de dualit\'e d\'efinissant
les potentiels magn\'etique $u$ et gravimagn\'etique $\chi$) se
r\'eduisent aux {\em \'equations d'Ernst} \cite{ernst}\ba
f\nabla^2\E &=& \nabla\E \cdot (\nabla\E + 2\ol{\psi}\psi)\,,\nn\\
f\nabla^2\psi &=& \nabla\psi \cdot (\nabla\E + 2\ol{\psi}\psi)\,,\nn\\
f^2R_{(3)ij} &=& \frac12(\E_{,(i} + 2\ol{\psi}\psi_{,(i}) (\E_{,j)}
+ 2\psi\ol{\psi}_{,j)}) - 2f\psi_{,(i}\ol{\psi}_{,j)} \ea ($_{,i}
\equiv \partial_i$, $(ij) \equiv \frac12(ij + ji)$), avec \be f =
{\rm Re}\,\E + \ol{\psi}\psi\,. \ee Ces \'equations sont \`a nouveau
celles d'un mod\`ele sigma coupl\'e \`a la gravitation \`a trois
dimensions, avec cette fois un espace cible \`a quatre dimensions.
L'ensemble de leurs invariances (manifestes et cach\'ees) forme le
groupe \`a 8 param\`etres $SU(2,1)$, l'espace cible \'etant le coset
\ba & \T & = SU(2,1)/S[U(1,1)\times U(1)]\,.
\\ ({\rm dim}:   & 4 & = \quad 8 \qquad - \qquad 4 \;) \nn
\ea Un point de ce coset peut \^etre repr\'esent\'e par la matrice
unitaire et unimodulaire \be \M(\E,\psi) =
f^{-1}\left(\begin{array}{ccc} 1 & \sqrt2\,\psi & \frac{i}2(\ol{\E}
- \E + 2\ol{\psi}\psi) \\ \sqrt2\,\ol{\psi} & -\frac12(\ol{\E} + \E
- 2\ol{\psi}\psi) & -i\sqrt2\,\E\ol{\psi} \\ \frac{i}2(\ol{\E} - \E
- 2\ol{\psi}\psi) & i\sqrt2\,\ol{\E}\psi & \E\ol{\E}
\end{array}\right)\,. \ee

\subsection{De Schwarzschild \`a Reissner-Nordstr\"{o}m}
Une repr\'esentation plus simple \`a utiliser est la
repr\'esentation de Kinnersley \cite{kin73}. On introduit trois
potentiels complexes $U$, $V$, $W$, reli\'es aux potentiels d'Ernst
par \be\lb{kinnersley} \E = \frac{U-W}{U+W}\,, \quad \psi =
\frac{V}{U+W}\,, \ee conduisant \`a \be f = {\rm Re}\,\E +
\ol{\psi}\psi = \frac{|U|^2 + |V|^2 -|W|^2}{|U+W|^2} \,. \ee La
redondance de cette param\'etrisation (trois potentiels complexes
$=$ six potentiels r\'eels, au lieu de quatre initialement) n'est
qu'apparente, les potentiels de Kinnersley n'\'etant d\'efinis par
(\ref{kinnersley}) qu'\`a une dilatation complexe \be\lb{scale}
(U,V,W) \to \xi(U,V,W) \quad (\xi \in C) \ee pr\`es. Le groupe
$SU(2,1)$ agit lin\'eairement sur le ``spineur'' $(U,V,W)$, laissant
la norme $|U|^2 + |V|^2 -|W|^2$ et la m\'etrique r\'eduite
$d\sigma^2 \equiv h_{ij}dx^idx^j$ invariantes.

La m\'etrique de Schwarzschild peut \^etre \'ecrite sous la forme
(\ref{st41}),\be ds^2 = - \frac{r-2m}r\,dt^2 +
\frac{r}{r-2m}\left[dr^2 + r(r-2m)d\Omega^2\right]\,, \ee conduisant
aux potentiels d'Ernst \be \E = f = \frac{r-2m}r\,, \quad \psi = 0
\ee (solution neutre des \'equations d'Einstein-Maxwell), puis aux
potentiels de Kinnersley  \be U = r-m\,, \quad V = 0\,, \quad W =
m\,, \ee avec \be |U|^2 + |V|^2 -|W|^2 = r(r-2m)\,. \ee

Les potentiels de Kinnersley \'etant d\'efinis \`a une dilatation
complexe pr\`es, la m\'etrique de Schwarzschild peut \^etre
repr\'esent\'ee de fa\c{c}on \'equivalente par le spineur \be
(\tilde{U},\tilde{V},\tilde{W}) = (1-m/r,\;0,\;m/r)\,, \ee
asymptotique au spineur de Minkowski $(1,0,0)$. Ce comportement
asymptotique est pr\'eserv\'e, d'une part par les transformations du
sous-groupe $U(1,1)$ m\'elangeant $V$ et $W$, d'autre part par les
transformations du sous-groupe $U(1)$ modifiant la phase de $U$, et
plus g\'en\'eralement par les produits de ces deux types de
transformations \`a une dilatation complexe (\ref{scale}) pr\`es,
soit par les transformations du sous-groupe d'isotropie \be H =
S[U(1)\times U(1,1)] \subset SU(2,1)\,. \ee

Les transformations de $H$ conservant le caract\`ere \'electriquement neutre
($V=0$) des solutions appartiennent \`a $S[U(1)\times U(1)]$ , et sont donc
\'equivalentes (voir note \ref{EinEM}) aux transformations (\ref{NUT})
engendrant une charge NUT. Les transformations de $SU(1,1)$
\be\lb{charge}
\left(\begin{array}{c} U' \\ V' \\ W' \end{array}\right) =
\left(\begin{array}{ccc} 1 & 0 & 0 \\ 0 & \cosh2\alpha & \sinh2\alpha \\
0 & \sinh2\alpha & \cosh2\alpha \end{array}\right)
\left(\begin{array}{c} U \\ V \\ W \end{array}\right)
\ee
engendrent des solutions charg\'ees. En d\'efinissant les nouveaux param\`etres
\be
m' \equiv m \cosh2\alpha\,, \quad e' \equiv m \sinh2\alpha \qquad (m'^2-e'^2
= m^2)\,,
\ee
on obtient \`a partir de (\ref{charge})
\ba
U'+W' &=& r-m+m\cosh2\alpha = r + r'_- = r'\,, \nn\\
U'-W' &=& r-m-m\cosh2\alpha = r' - 2m'\,,\\
V' &=& e'\,,
\ea
qui conduisent \`a la solution
\ba
ds'^2 &=& \frac{(r'-r'_+)(r'-r'_-)}{r'^2}dt^2  + \frac{r'^2}
{(r'-r'_+)(r'-r'_-)}\left[dr'^2 + (r'-r'_+)(r'-r'_-)d\Omega^2\right]\,, \nn\\
\psi' & = & \frac{e'}{r'}\,, \ea o\`u l'on reconnait la solution de
Reissner-Nordstr\"{o}m (\ref{RN}) (les param\`etres $r'_{\pm}$ sont
reli\'es \`a la masse de Schwarzschild et au param\`etre $\alpha$ de
la transformation par $r'_+ = 2m\cosh^2\alpha$, $r'_- =
2m\sinh^2\alpha$).

\subsection{De Schwarzschild \`a Kerr}

Les sections spatiales de la m\'etrique de Schwarzschild sont \`a
sym\'etrie sph\'erique, tandis que celles de la m\'etrique de Kerr
sont seulement \`a sym\'etrie axiale. Donc, contrairement \`a
l'exemple pr\'ec\'edent, il n'est pas possible de transformer la
m\'etrique de Schwarzschild en celle de Kerr par une transformation
du groupe $G=SL(2,R)$ (ou $SU(2,1)$), ces transformations
pr\'eservant la m\'etrique r\'eduite \`a trois dimensions $h_{ij}$.
Pourtant, le travail de Geroch \cite{geroch} montre qu'il doit
exister une relation entre ces deux solutions des \'equations
d'Einstein.

\subsubsection{Groupe de Geroch}

Schwarzschild et Kerr ont en commun d'\^etre des solutions
stationnaires \`a sym\'etrie axiale des \'equations d'Einstein,
c'est-\`a-dire qu'elles ont deux vecteurs de Killing commutant enre
eux, $\partial_t$ et $\partial_{\varphi}$. Geroch a montr\'e qu'on
peut combiner \`a l'infini des transformations infinit\'esimales du
groupe $G$ (associ\'e \`a la r\'eduction dimensionnelle par rapport
\`a un vecteur de Killing donn\'e) avec des transformations
lin\'eaires infinit\'esimales dans l'espace \`a deux dimensions
engendr\'e par $\partial_t$ et $\partial_{\varphi}$, conduisant
ainsi \`a une structure appel\'ee (improprement) ``groupe'' de
Geroch.

L'existence de cette structure permet en principe de g\'en\'erer
toutes les solutions stationnaires axisym\'etriques de E4 (ou de
EM4) par des transformations du groupe. Il en r\'esulte que E4 ou
EM4 avec deux vecteurs de Killing sont des th\'eories compl\`etement
int\'egrables. Les m\'ethodes pratiques de construction de solutions
exploitant cette propri\'et\'e sont les m\'ethodes de transformation
de diffusion inverse (inverse scattering transform) \cite{BZ79}.

Que se passe-t-il si on consid\`ere une transformation de Geroch,
non plus infinit\'esimale, mais finie? Les transformations du groupe
$GL(2,R)$ agissant sur l'espace ($\partial_t$, $\partial_{\varphi}$)
sont a priori \be \left(\begin{array}{c} d\varphi' \\ dt'
\end{array}\right) = \left(\begin{array}{cc} \alpha & \beta \\
\gamma & \delta \end{array}\right) \left(\begin{array}{c} d\varphi
\\ dt \end{array}\right)\,. \ee Mais le fait que
$\partial_{\varphi}$ engendre non pas des translations, mais des
rotations (ses orbites sont ferm\'ees) conduit \`a des restrictions.
D'abord, la p\'eriodicit\'e de la variable angulaire $\varphi$ est
pr\'eserv\'ee \`a condition que $\alpha = 1$. Ensuite, la nouvelle
variable temps $t' = \gamma\varphi + \delta t$ doit \^etre
p\'eriodiquement identifi\'ee (modulo $2\pi$), conduisant \`a
l'existence de courbes ferm\'ees de genre temps violant la
causalit\'e, \`a moins que $\gamma = 0$. Apr\`es ces restrictions,
il ne reste plus que des transformations \`a deux param\`etres qui
s'\'ecrivent (apr\`es red\'efinition de ces param\`etres)

\be\lb{ROg}
{\cal R}(\Omega,\gamma)\quad \left\vert \begin{array}{ll} d\varphi' =
d\varphi - \Omega dt \\ dt' = \gamma^{-1}dt \end{array}\right. \,.
\ee
Ces transformations consistent donc essentiellement dans le passage \`a un
rep\`ere tournant \`a la vitesse angulaire $\Omega$, combin\'e
\'eventuellement avec un changement d'\'echelle des temps.

Le probl\`eme, c'est que le passage \`a un rep\`ere tournant modifie
le comportement asymptotique. En m\'ecanique non relativiste, le
passager du rep\`ere tournant subit une ``force centrifuge'', certes
fictive (elle peut \^etre annul\'ee localement en passant \`a un
rep\`ere acc\'el\'er\'e), mais dont les effets sont pourtant
sensibles. En relativit\'e g\'en\'erale, la transformation de
coordonn\'ees (\ref{ROg}) agissant sur la m\'etrique de Minkowski
\'ecrite en coordonn\'ees sph\'eriques conduit \`a la m\'etrique \be
ds^{'2} = -\gamma^2dt^{'2} + dr^2 + r^2[d\theta^2 + \sin^2\theta
(d\varphi' + \Omega\gamma dt')^2]\,. \ee Cette m\'etrique est
toujours plate (puisqu'on a simplement effectu\'e une transformation
globale de coordonn\'ees), mais \'evidemment non asymptotiquement
minkowskienne, puisque \be -g'_{tt} = \gamma^2(1 - \Omega^2 r^2) \ee
change de signe au-del\`a du cercle $r = \Omega^{-1}$ dont la
vitesse de rotation est \'egale \`a celle de la lumi\`ere!

Malgr\'e ce probl\`eme, il est possible de construire, dans le cadre de la
th\'eorie d'Einstein-Maxwell, des transformations de Geroch finies qui
pr\'eservent le caract\`ere asymptotiquement minkowskien des solutions. La
solution \cite{kerr} consiste \`a combiner le changement de rep\`ere
${\cal R}(\Omega,\gamma)$ avec une transformation $\Pi \in G = SU(2,1)$
modifiant aussi le comportement asymptotique.

\subsubsection{De Schwarzschild \`a Bertotti-Robinson}

R\'e\'ecrivons la m\'etrique de Schwarzschild (apr\`es translation
$r \to r+m_0$ de la coordonn\'ee radiale): \be ds^2 = -
\frac{r-m_0}{r+m_0}dt^2 + \frac{r+m_0}{r-m_0}[dr^2 + (r^2 - m_0^2)
d\Omega^2]\,, \ee et la m\'etrique de Bertotti-Robinson ($AdS_2
\times S^2$) \be d\hat{s}^2 = \left(1 - \frac{r^2}{m_0^2}\right)dt^2
+ \left(\frac{r^2}{m_0^2} -1\right)^{-1}[dr^2 + (r^2 -
m_0^2)d\Omega^2]\,. \ee Ces deux 4-m\'etriques ont m\^eme m\'etrique
r\'eduite \`a trois dimensions (entre crochets). Il existe donc une
transformation $\Pi \in SU(2,1)$ qui relient entre elles les
solutions de Schwarzschild (asymptotiquement minkowskienne) et de
Bertotti-Robinson (non asymptotiquement minkowskienne). Cette
transformation est l'involution ($\Pi^{-1} = \Pi$) \be \Pi\quad:
\qquad U \longleftrightarrow V\,. \ee En effet, son action sur la
solution de Schwarzschild conduit \`a \ba
&&(U=r,\;V=0,\;W=m_0) \longrightarrow (\hat{U}=0,\;\hat{V}=r,\;\hat{W}=m_0) \nn\\
&&\hat{f} = \frac{|\hat{U}|^2 + |\hat{V}|^2 -|\hat{W}|^2}{|\hat{U}+\hat{W}|
^2} = \frac{r^2-m_0^2}{m_0^2}\,, \quad \hat{\psi} = \frac{\hat{V}}{\hat{U}
+ \hat{W}} = \frac{r}{m_0}\,,
\ea
qui caract\'erisent la solution de Bertotti-Robinson.

Dans le langage des potentiels d'Ernst, la transformation $\Pi$ agissant sur
une solution de potentiels ($\E,\;\psi$) conduit \`a une solution de
potentiels ($\hat{\E},\;\hat{\psi}$) avec
\be
\hat{\E} = \frac{-1+\E+2\psi}{1-\E+2\psi}\,, \quad \hat{\psi} =
\frac{1+\E}{1-\E+2\psi}\,.
\ee
Elle transforme une solution neutre $\psi = 0$ ($V=0$) des \'equations
d'Einstein-Maxwell en une solution avec $\hat{\E}=-1$ ($\hat{U}=0$), et une
solution \'electrostatique ($\E$ et $\psi$ r\'eels) asymptotiquement
minkowskienne en une solution \'electrostatique ($\hat{\E}$, $\hat{\psi}$)
asymptotiquement Bertotti-Robinson.

Effectuons maintenant sur cette derni\`ere solution ``chapeaut\'ee''
la transformation ${\cal R}(\Omega,\gamma)$ (passage \`a un rep\`ere
tournant). Comme nous allons le voir en d\'etail, cette
transformation m\'elange le $AdS_2$ et le $S^2$ sans modifier
fondamentalement le caract\`ere asymptotiquement Bertotti-Robinson
de la nouvelle solution ($\hat{\E'}$, $\hat{\psi'}$). Par contre,
cette derni\`ere n'est plus \'electrostatique, le passage \`a un
rep\`ere tournant induisant un champ magn\'etique (partie imaginaire
de $\hat{\psi'}$). L'application \`a ($\hat{\E'}$, $\hat{\psi'}$) de
la transformation inverse $\Pi^{-1} = \Pi$ va donc conduire \`a une
solution ($\E',\psi'$) asymptotiquement plate, avec en g\'en\'eral
un moment dipolaire magn\'etique (qui pourra \^etre annul\'e par un
choix judicieux des param\`etres $\Omega$ et $\gamma$) et un moment
dipolaire gravimagn\'etique, c'est-\`a-dire un moment angulaire.
Donc, au final, la transformation combin\'ee finie de Geroch \be
\Sigma = \Pi^{-1}{\cal R}\Pi \ee g\'en\`ere une solution en rotation
\`a partir d'une solution statique!

\subsubsection{De Bertotti-Robinson \`a Kerr}
Revenons au cas particulier de la solution de Bertotti-Robinson (la
transform\'ee par $\Pi$ de la solution de Schwarzschild). Par
commodit\'e, r\'e\'ecrivons-l\`a en termes de la coordonn\'ee
temporelle sans dimension $\tau = t/m_0$ et des coordonn\'ees
sph\'ero\"{\i}dales (\'egalement sans dimension) $x = r/m_0$, $y =
\cos\theta$: \ba d\hat{s}^2  &=&  m_0^2\left[(1-x^2)d\tau^2  +
\frac{dx^2}{x^2-1}
+ \frac{dy^2}{1-y^2} + (1-y^2) d\varphi^2 \right] \,,   \nn \\
\hat{A} &=& m_0x\,d\tau\,.
\ea
L'action sur cette solution de la transformation
\be
{\cal R}(\Omega,\gamma)\quad \left\vert \begin{array}{ll} d\varphi =
d\varphi' + \gamma\eta\,d\tau' \\ d\tau = \gamma\,d\tau' \end{array}\right.
\quad (\eta = m_0\Omega)
\ee
conduit \`a la ``nouvelle'' solution \footnote{En fait, il s'agit bien s\^ur
toujours de l'espace-temps de Bertotti-Robinson, repr\'esent\'e dans un
nouveau syst\`eme de coordonn\'ees.}
\ba\lb{BR2}
d\hat{s}'^2  &=&  m_0^2 \left[ -\gamma^2(x^2+\eta^2y^2-(1+\eta^2))
\left(d\tau' -\frac
{\eta(1-y^2)}{\gamma(x^2+\eta^2y^2-(1+\eta^2))}\,d\varphi'\right)^2
\right. \nn \\ && \left. + \frac{dx^2}{x^2-1}  + \frac{dy^2}{1-y^2}
+ \frac{(x^2-1)(1-y^2)}{x^2+\eta^2y^2-(1+\eta^2)}\,d\varphi'^2\right] \nn\\
\hat{A}' &=& m_0\gamma x\,d\tau'\,, \ea qui a le m\^eme comportement
asymptotique pour $x \to \infty$ (aux facteurs $\gamma$ et
$1+\eta^2$ pr\`es). Le champ magn\'etique induit, donn\'e par \be
\hat{F}^{'x\varphi} =
\frac{\partial\varphi'}{\partial\tau}\hat{F}^{'x\tau} =
\frac{\eta}{m_0^4}\,, \ee conduit (en utilisant (\ref{magn}), o\`u
$\sqrt{|h|}/f = m_0^4\gamma$), et (\ref{vu})) au potentiel scalaire
magn\'etique $u' = -\eta\gamma y$. Apr\`es calcul du potentiel
scalaire de twist par r\'esolution de l'\'equation de dualit\'e
(\ref{chi}), on obtient par (\ref{Epsi}) les potentiels d'Ernst
transform\'es \be \hat{\cal E'} =  - (p^2+q^2)\,, \quad  \hat{\psi'}
= px - iqy \qquad (p \equiv \gamma\,, q \equiv \eta\gamma)\,. \ee En
particulier, le choix de la relation entre les param\`etres \be
p^2+q^2=1 \ee ($\gamma = (1 + \eta^2)^{-1/2}$) assure que $\hat{\cal
E}' = -1$, et donc que la solution transform\'ee
$(\E',\psi')=\Pi^{-1}(\hat{\E'}$, $\hat{\psi'})$ est neutre. Les
potentiels d'Ernst de cette solution finale \be {\cal E}' = \frac{px
- iqy - 1}{px - iqy + 1}\,, \quad   \psi' = 0 \ee sont ceux de la
solution de Kerr de param\`etres $m = m_0p$, $a = m_0q$, exprim\'es
en coordonn\'ees sph\'ero\"{\i}dales \cite{ernst}.

\subsection{Gravitation \`a cinq dimensions}

\subsubsection{Le mod\`ele sigma}
Consid\'erons la gravitation d'Einstein sans source \`a cinq dimensions (E5).
Dans le cas o\`u la 5-m\'etrique admet un vecteur de Killing $\partial_5$
de genre espace, la th\'eorie se r\'eduit \`a une th\'eorie effective \`a
quatre dimensions (une g\'en\'eralisation dilatonique de la th\'eorie
d'Einstein-Maxwell) appel\'ee th\'eorie de Kaluza-Klein. Les solutions
stationnaires de cette th\'eorie admettent elles-m\^emes un vecteur de Killing
$\partial_4$ de genre temps, permettant de r\'eduire \`a nouveau \`a trois
dimensions. Il est plus \'el\'egant de r\'eduire directement E5 avec deux
vecteurs de Killing (un de genre temps, un de genre espace) de cinq \`a trois
dimensions en conservant explicitement la covariance sous $GL(2,R)$
\cite{maison,karim}:
\begin{equation}\label{st5}
ds^2 = \lambda_{ab}(dx^a + \A_i^adx^i)(dx^b
+ \A_j^bdx^j) + \tau^{-1}h_{ij}dx^idx^j\,,
\end{equation}
avec $a,b=4,5$, $i,j=1,2,3$, et $\tau \equiv -$det$\lambda$. La
r\'eduction des \'equations d' Einstein \`a cinq dimensions
g\'en\'eralise celle effectu\'ee \`a quatre dimensions, l'\'equation
de dualit\'e (\ref{om2}) \'etant remplac\'ee par l'\'equation
2-covariante
\begin{equation}\label{duale5}
\tau\lambda_{ab}\nabla\wedge\vec{\A}^b = \nabla\omega_a\,.
\end{equation}
Il en r\'esulte un mod\`ele sigma auto-gravitant pour un espace
cible \`a cinq dimensions (trois potentiels ``gravi\'electriques''
$\lambda_{ab}$ et deux potentiels ``gravimagn\'etiques''
$\omega_a$). Neugebauer \cite{neuge}, puis Maison \cite{maison} ont
montr\'e que cet espace cible admet huit vecteurs de Killing,
engendrant le groupe $SL(3,R)$. Ce groupe agit bilin\'eairement sur
la matrice repr\'esentative $3\times3$ sym\'etrique et unimodulaire
\begin{equation}\label{maimat}
\chi = \left(\begin{array}{cc}
\lambda - \tau^{-1}\omega\omega^T &
\tau^{-1}\omega \\ \tau^{-1}\omega^T & -\tau^{-1}
\end{array}\right)\,,
\end{equation}
o\`u $\lambda$ est un bloc $2\times2$, et $\omega$ est un 2-vecteur
colonne. De fa\c{c}on analogue au cas de E4, les \'equations r\'eduites
\begin{eqnarray}
\nabla(\chi^{-1}\nabla\chi) &=& 0\,, \label{divJ5} \\
R_{ij}(h) &=& \frac14{\rm Tr}(\chi^{-1}\partial_i\chi
\chi^{-1}\partial_j\chi)\,, \label{Rij5}
\end{eqnarray}
sont manifestement invariantes sous $G = SL(3,R)$.

Les solutions int\'eressantes du point de vue de la th\'eorie de
Kaluza-Klein sont asymptotiques \`a un cylindre d'espace-temps, produit
direct d'un cercle (la cinqui\`eme dimension compactifi\'ee) par Minkowski \`a
quatre dimensions. Dans la terminologie moderne, les trous noirs de ce type
sont appel\'es cordes noires (``black strings''). Dans le cas d'un tel
comportement asymptotique, la matrice de Maison (\ref{maimat}) est asymptote
\`a
\begin{equation}\label{etabs}
\chi_{\infty} = \eta_{BS} = \left(\begin{array}{ccc}
-1 & 0 & 0 \\ 0 & 1 & 0 \\ 0 & 0 & -1
\end{array}\right)\,.
\end{equation}
Cette matrice \'etant pr\'eserv\'ee par le groupe d'isotropie $H =
SO(2,1)$, l'espace cible de E5 est $SL(3,R)/SO(2,1)$. Cette
d\'emarche se g\'en\'eralise simplement au cas de la th\'eorie
d'Einstein \`a $4+p$ dimensions avec $(p+1)$ vecteurs de Killing
(E(4+p)), dont l'espace cible est $SL(2+p,R)/SO(2,p)$.

Une des premi\`eres applications fructueuses de ce mod\`ele sigma
bas\'e sur $SL(3,R)$ a \'et\'e la construction de solutions de
Kaluza-Klein multi-centres \cite{spat}, qui a servi de mod\`ele pour
le sch\'ema g\'en\'eral expos\'e au 3.4. Plus tard, Rasheed
\cite{rasheed} a g\'en\'er\'e des solutions cordes noires en
rotation \`a partir de la solution de Kerr, par la transformation
\begin{equation}
\chi = P^T \chi_K P\,,
\end{equation}
o\`u $\chi_K$ est la matrice de Maison pour la corde noire de Kerr
\begin{equation}\label{bsk}
ds_{(5)}^2 = (dx^5)^2 + ds_K^2\,,
\end{equation}
($ds_K^2$ \'etant la m\'etrique de Kerr \`a quatre dimensions), et $P$
une matrice constante de $SO(2,1)$, contrainte de fa\c{c}on \`a ce que la
charge NUT de la solution r\'esultante s'annulle.

\subsubsection{Des cordes noires aux trous noirs}
A c\^ot\'e des solutions cordes noires, dont l'horizon a la
topologie $S^1\times S^2$, la th\'eorie d'Einstein \`a cinq
dimensions a aussi des solutions trous noirs (``black holes"), dont
l'horizon a la topologie $S^3$. Les solutions stationnaires trous
noirs \`a cinq dimensions, g\'en\'eralisant la solution de Kerr \`a
quatre dimensions, ont \'et\'e construites par Myers et Perry
\cite{mype}. Le groupe de rotation \`a quatre dimensions d'espace
admettant deux g\'en\'erateurs commutant entre eux, ces solutions
sont donc param\'etr\'ees par trois ``nombres quantiques'' $m$
(masse), $a_+$ et $a_-$ (reli\'es aux deux moments angulaires).

La solution trou noir statique \`a cinq dimensions, obtenue
initialement par Tangherlini \cite{tang}, s'\'ecrit \be\label{tan1}
ds_T^2 = -\bigg(1-\frac{\mu}{\rho^2}\bigg)dt^2 +
\bigg(1-\frac{\mu}{\rho^2}\bigg)^{-1}d\rho^2 + \rho^2d\Omega_3^2\,,
\end{equation}
o\`u
\begin{equation}
 d\Omega_3^2 =
\frac14\bigg[(d\eta-\cos\theta d\varphi)^2 + d\theta^2 +
\sin^2\theta d\varphi^2\bigg]
\end{equation}
est la m\'etrique de la trois-sph\`ere. La transformation de
coordonn\'ees $\rho^2 = 4mr$, $\eta = x^5/m$, avec $m^2 = \mu/8$,
permet de r\'e\'ecrire (\ref{tan1}) sous la forme \ba\label{tan2}
ds_T^2 &=& -\frac{r-2m}{r}\,dt^2 + \frac{r}{m}(dx^5-m\cos\theta
d\varphi)^2 \nn\\ && + \frac{m}{r-2m}\bigg[dr^2 +
r(r-2m)\bigg(d\theta^2 + \sin^2\theta d\varphi^2\bigg)\bigg]\,. \ea
Cette 5-m\'etrique de Tangherlini a la m\^eme m\'etrique r\'eduite
\`a trois dimensions que la corde noire de Schwarzschild,
\begin{equation}\label{sch}
ds_S^2 = -\frac{r-2m}{r}\,dt^2 + (dx^5)^2 + \frac{r}{r-2m}\bigg[dr^2
+ r(r-2m)\bigg(d\theta^2 + \sin^2\theta d\varphi^2\bigg)\bigg]\,,
\end{equation}
donc leurs deux matrices de Maison doivent \^etre reli\'ees par une
transformation de $SL(3,R)$
\begin{equation}\label{transfST}
\chi_T = P_{ST}^T\chi_SP_{ST}\,.
\end{equation}
La matrice de transformation $P_{ST}$ a \'et\'e d\'etermin\'ee dans
\cite{newdil2}. Cette matrice n'appartient pas au sous-groupe
$SO(2,1)$, parce que les m\'etriques (\ref{sch}) et (\ref{tan2}) ont
des comportements asymptotiques diff\'erents, ce qui fait que la
matrice $\chi_T$ tend pour $r \to \infty$ vers une matrice constante
$\eta_{BH}$ (donn\'ee plus loin en (\ref{etabh})) diff\'erente de la
matrice $\eta_{BS}$ de (\ref{etabs}).

Toutes les 5-m\'etriques du type ``corde noire'' ayant le m\^eme
comportement asymptotique que (\ref{sch}), et toutes les
5-m\'etriques du type ``trou noir'' ayant le m\^eme comportement
asymptotique que (\ref{tan2}), on s'attend \`a ce que de fa\c{c}on
g\'en\'erale la transformation (\ref{transfST}) transforme des
cordes noires en trous noirs:
\begin{equation}\label{bsbh}
\chi_{BH} = P_{ST}^T\chi_{BS}P_{ST}\,.
\end{equation}
En effet, nous avons trouv\'e dans \cite{newdil2,cedric} que
l'action de cette transformation sur la corde noire de Kerr
(\ref{bsk}) conduit au trou noir de Myers-Perry avec des moments
angulaires oppos\'es, $a_+ = -a_-$. Nous avons aussi remarqu\'e que,
dans le cas g\'en\'eral,  la 3-m\'etrique r\'eduite du trou noir de
Myers-Perry avec des param\`etres $a_+$ et $a_-$ quelconques
co\"{\i}ncide \`a nouveau avec celle de la corde noire de Kerr. Dans
le cas particulier du trou noir de Myers-Perry avec des moments
angulaires {\em \'egaux}, $a_+ = a_-$, la m\'etrique \`a quatre
dimensions est statique (dyonique), et la 3-m\'etrique r\'eduite est
la m\^eme que celle de la corde noire de Schwarzschild. Mais la
port\'ee de cette observation nous a \'echapp\'e \`a l'\'epoque.

\subsubsection{De Tangherlini \`a Myers-Perry}

Giusto et Saxena \cite{giusax} ont remarqu\'e ind\'ependemment que
la matrice de Maison asymptotique n'est pas la m\^eme pour les
cordes noires et pour les trous noirs. Dans le cas de la m\'etrique
de Tangherlini (\ref{tan2}), $\omega_5=r/m + b$ ($b$ constante
d'int\'egration), donc $\chi_{45}=\tau^{-1}\omega_5 \to 1$, tandis
que $\chi_{55}$ tend vers une limite finie qu'on peut annuler par un
choix appropri\'e de la constante $b$. Dans ce cas,
\begin{equation}\label{etabh}
\eta_{BH} = \left(\begin{array}{ccc} -1 & 0 & 0 \\ 0 & 0 & 1
\\ 0 & 1 & 0
\end{array}\right)\,.
\end{equation}
Les deux matrices $\eta_{BH}$ et $\eta_{BS}$ sont reli\'ees par la
transformation (\ref{transfST}) avec la matrice de transformation
\begin{equation}\label{P3ST}
P_{ST} = \left(\begin{array}{ccc} 1 & 0 & 0 \\ 0 & 1/\sqrt2 &
1/\sqrt2 \\ 0 & -1/\sqrt2 & 1/\sqrt2
\end{array}\right)\,.
\end{equation}

Le groupe d'isotropie $SO(2,1)$ pr\'eservant $\eta_{BH}$ est
engendr\'e par trois transformations $M_{\alpha}$, $M_{\beta}$ et
$M_{\gamma}$, ces deux derni\`eres triviales (ce sont des
transformations de jauge g\'en\'eralis\'ees). Giusto et Saxena ont
montr\'e que la transformation $M_{\alpha}$ permet d'engendrer le
trou noir de Myers-Perry \`a partir du trou noir de Tangherlini en
trois \'etapes:
\begin{enumerate}
\item L'action de $M_{\alpha}$ sur la m\'etrique de Tangherlini (\ref{tan1})
(o\`u la m\'etrique de $S^3$ est \'ecrite sous une forme
sym\'etrique en ($\eta$, $\varphi$))
\begin{equation}\label{tan3}
ds_T^2 = -\bigg(1-\frac{\mu}{\rho^2}\bigg)dt^2 +
\bigg(1-\frac{\mu}{\rho^2}\bigg)^{-1}d\rho^2 + \frac{\rho^2}4
(d\theta^2 + d\eta^2 + d\varphi^2 - 2\cos\theta d\eta d\varphi),
\end{equation}
r\'eduite par rapport aux vecteurs de Killing $\partial_t$ et
$\partial_{\eta}$, conduit \`a la m\'etrique de Myers-Perry avec
deux moments angulaires \'egaux, $a_+ = a_-$.
\item L'\'echange ("flip") des angles $\eta\leftrightarrow \varphi$ revient \`a
r\'eduire la m\'etrique de Myers-Perry avec $a_+ = a_-$ par rapport
aux vecteurs de Killing $\partial_t$ et $\partial_{\varphi}$ au lieu
de $\partial_t$ et $\partial_{\eta}$ (une transformation finie de
Geroch!) ou, de fa\c{c}on \'equivalente, \`a remplacer la m\'etrique
de Myers-Perry avec deux moments angulaires \'egaux par la
m\'etrique de Myers-Perry avec deux moments angulaires oppos\'es.
\item L'action sur cette derni\`ere d'une nouvelle transformation
$M_{\alpha'}$ conduit \`a une m\'etrique de Myers-Perry g\'en\'erique
avec $a_+ \neq \pm a_-$.
\end{enumerate}

\subsection{Coda}
En cinq dimensions, les topologies possibles pour les objets noirs
sont plus vari\'ees qu'en quatre dimensions, et parall\`ellement les
mod\`eles sigma offrent de nouvelles possibilit\'es pour transformer
ces objets noirs entre eux. Ainsi, la combinaison entre les
transformations $P_{ST}$ du 3.8.2 et $GS$ (Giusto-Saxena) du 3.8.3
fournit une autre possibilit\'e de transformer la m\'etrique de
Schwarzschild en la m\'etrique de Kerr, cette fois dans le cadre du
groupe $SL(3,R)$, par la suite des op\'erations:
$$\begin{array}{ccccccc}S & \longrightarrow & T & \longrightarrow &
MP & \longrightarrow & K \\ & P_{ST} && GS && P_{ST}^{-1} &
\end{array}$$

A c\^ot\'e des cordes noires et des trous noirs, il existe aussi en
cinq dimensions des anneaux noirs (``black rings"), qu'on peut
visualiser comme des cordes noires ferm\'ees. Ceux-ci ne sont
r\'eguliers que s'ils poss\`edent au moins un moment angulaire
\cite{ER02b}. Des anneaux r\'eguliers avec deux moments angulaires
ont \'et\'e construits par la m\'ethode de transformation de
diffusion inverse \cite{pomsen}. En principe, ils devraient \^etre
reli\'es \`a ceux de \cite{ER02b} par une transformation de
Giusto-Saxena mais, curieusement, ceci n'a pas encore \'et\'e
v\'erifi\'e \cite{bring}.

La th\'eorie d'Einstein-Maxwell \`a quatre dimensions peut \^etre
consid\'er\'ee comme le secteur bosonique de la supergravit\'e avec
${\cal N} = 2$ \`a quatre dimensions. Sa g\'en\'eralisation
naturelle \`a cinq dimensions est le secteur bosonique de la
supergravit\'e minimale \`a cinq dimensions, c'est-\`a-dire la
th\'eorie d'Einstein-Maxwell avec un terme de Chern-Simons (EM5). Le
groupe d'invariance de cette th\'eorie r\'eduite \`a trois
dimensions est le groupe de Lie exceptionnel $G2$, qui a \'et\'e
utilis\'e pour engendrer des trous noirs et des anneaux noirs
charg\'es \cite{g2}. Dans le cadre de ce groupe, on peut \'egalement
g\'en\'eraliser la m\'ethode du 3.7 pour ajouter un ou deux moments
angulaires \`a un trou noir statique, la solution interm\'ediaire
``chapeaut\'ee'' \'etant cette fois l'\'equivalent pour EM5 de la
solution de Bertotti-Robinson ($AdS_2\times S^3$ ou $AdS_3\times
S^2$) \cite{bremen}, mais l'application de cette m\'ethode aux
anneaux noirs conduit \`a des r\'esultats difficiles \`a
interpr\'eter.

Enfin, de m\^eme qu'il est possible de transformer des cordes noires
en trous noirs, serait-il possible de transformer des trous noirs en
anneaux noirs, en passant \`a nouveau par l'interm\'ediaire de
transformations de coordonn\'ees sur la solution de
Bertotti-Robinson \`a cinq dimensions? La question m\'erite d'\^etre
pos\'ee.

\end{document}